\documentclass[lettersize,journal]{IEEEtran} 

\usepackage[utf8]{inputenc}
\usepackage[T1]{fontenc}
\usepackage[english]{babel}

\usepackage{amsmath,amsfonts}
\usepackage{array}
\usepackage{textcomp} 
\usepackage{nth} 
\usepackage{stfloats}
\usepackage{url}
\usepackage{verbatim}
\usepackage{graphicx}
\usepackage{cite}

\usepackage{minted} 
\usepackage{algorithm} 
\usepackage[noend]{algpseudocode} 

\usepackage{booktabs} 
\usepackage{tabularx} 
\usepackage{multirow} 
\usepackage{multicol} 
\usepackage{colortbl} 
\usepackage[para]{threeparttable}
\usepackage{changepage}
\usepackage{soul} 
\usepackage{adjustbox}
\usepackage{float}

\usepackage[table]{xcolor}
\usepackage{cellspace}
\usepackage{svg} 
\usepackage{orcidlink}
\usepackage{placeins} 
\usepackage{tikz}
\usepackage{amssymb}
\usepackage[font=small,labelfont=bf]{caption}
\usepackage{subcaption}

\usepackage[capitalise]{cleveref} 

\usepackage{siunitx} 
\usepackage{xfrac} 
\usepackage{xspace} 
\usepackage[inline]{enumitem} 
\usepackage{footnote}
\usepackage{textgreek}
\usepackage[nolist]{acronym} 
\usepackage{lipsum} 

\def\BibTeX{{\rm B\kern-.05em{\sc i\kern-.025em b}\kern-.08em
    T\kern-.1667em\lower.7ex\hbox{E}\kern-.125emX}}

\algrenewcommand\alglinenumber[1]{\scriptsize #1:}

\algrenewcommand\algorithmicloop{\textbf{loop}\xspace}
\algnewcommand{\algorithicgoto}{\textbf{goto}}%
\algnewcommand{\Goto}[1]{\algorithicgoto~\ref{#1}}%
\algnewcommand{\algorithmicnot}{\textbf{not}}
\algdef{SE}[IF]{IfNot}{EndIf}[1]{\algorithmicif\ \algorithmicnot\ #1\ \algorithmicthen}{\algorithmicend\ \algorithmicif}%
\makeatletter
\ifthenelse{\equal{\ALG@noend}{t}}%
  {\algtext*{EndIf}}
  {}%
\makeatother

\definecolor{lightgray}{gray}{0.98}
\definecolor{cpu}{HTML}{434343}
\definecolor{carus}{HTML}{007480}
\definecolor{caesar}{HTML}{6d1a36}
\begin{acronym}
    \acro{flsl}[FLSL]{Feedforward Leakage Self-suppression Logic}
    \acro{emv}[EMV]{energy-minimum voltage}
\end{acronym}

\newcommand{\fdsoi}{\ac{fdsoi}\xspace}
\newcommand{\flsl}{\ac{flsl}\xspace}
\newcommand{\dlsl}{\ac{dlsl}\xspace}
\newcommand{\hvt}{\ac{hvt}\xspace}
\newcommand{\uhvt}{\ac{uhvt}\xspace}
\newcommand{\ehvt}{\ac{ehvt}\xspace}
\newcommand{\lvt}{\ac{lvt}\xspace}
\newcommand{\fbb}{\ac{fbb}\xspace}
\newcommand{\rbb}{\ac{rbb}\xspace}
\newcommand{\ao}{\ac{ao}\xspace}
\newcommand{\pvt}{\ac{pvt}\xspace}
\newcommand{\ldp}{\ac{ldp}\xspace}
\newcommand{\aes}{\ac{aes}\xspace}

\begin{document}

\newcommand{\flsltitle}{Enabling Ultra-Low-Power Always-On Feedforward Leakage Suppression Logic Circuits with FDSOI }
\title{\flsltitle}

\author{
    Clément Choné\orcidlink{0009-0000-2625-9019}, Leslie Xu\orcidlink{0009-0005-7529-567X}, Filippo Quadri\orcidlink{0009-0004-7047-8445}, Pasquale Davide Schiavone\orcidlink{0000-0003-2931-0435}, Alexandre Levisse\orcidlink{0000-0002-8984-9793}, Jean-Luc Naguel\orcidlink{}, David Atienza\orcidlink{0000-0001-9536-4947}~\IEEEmembership{Fellow, IEEE}, and Andreas Burg \orcidlink{}%

    \thanks{C. Choné is with the Embedded Systems Laboratory (ESL) and with the Telecommunications Circuits Laboratory (TCL), EPFL, 1015 Lausanne, Switzerland (e-mail: clement.chone@epfl.ch). L. Xu is with the Swiss Center for Microelectronics and Microtechnology (CSEM), 2000 Neuchâtel and with the Telecommunications Circuits Laboratory (TCL), EPFL (email: shijie.xu@csem.ch). J-L. Naguel is with the Swiss Center for Microelectronics and Microtechnology (CSEM) (e-mail: jean-luc.nagel@csem.ch). F. Quadri, P. D. Schiavone, A. Levisse and D. Atienza are with the Embedded Systems Laboratory (ESL), EPFL (e-mail: filippo.quadri@epfl.ch; davide.schiavone@epfl.ch; alexandre.levisse@epfl.ch; david.atienza@epfl.ch). A. Burg is with the Telecommunications Circuits Laboratory (TCL), EPFL, 1015 Lausanne, Switzerland (e-mail: andreas.burg@epfl.ch).}%

}

\markboth{IEEE Transactions on Very Large Scale Integration Systems,~Vol.~XX, No.~XX, XX~2026}%
{Choné\MakeLowercase{\textit{et al.}}: \flsltitle}


\maketitle


\begin{abstract} 
The growing deployment of real-time applications on wearable and Internet of Things (IoT) edge devices has intensified the need for energy-efficient, high-performance systems that meet stringent timing and energy constraints. Due to their favorable energy–delay product, \fdsoi technologies have become a compelling choice for designing such systems. In addition to the technology, events-driven architectures leverage the sparsity of real-time to further improve system energy efficiency by employing an \ao domain to monitor inputs and activate a high-performance (HP) domain only when relevant events occur. However, for low-duty-cycle applications or when input signals vary slowly and reliable wake-up detection incurs non-negligible complexity, the energy bottleneck shifts toward the \ao domain, where leakage power dominates overall consumption.
To mitigate this issue, \ao circuits are typically implemented using \hvt or \uhvt transistors, thereby avoiding sub- and near-threshold operation, which is highly sensitive to \pvt variations. Nonetheless, given their relatively small size compared to the main system, \ao subsystems are well-suited to alternative logic styles that trade off area for power efficiency. In this context, \flsl has recently emerged as a promising candidate, offering reduced leakage compared to conventional CMOS while maintaining higher operating speed than prior leakage reduction techniques.
However, previous studies report a significant degradation in FLSL leakage performance in technology nodes below \SI{90}{\textbf{\nano\meter}}, primarily due to increased gate and junction leakage currents. \fdsoi technology, with its ability to effectively suppress junction leakage, provides an opportunity to overcome this limitation and restore FLSL efficiency in advanced nodes.
Therefore, we demonstrate in this work that \flsl implemented in a \SI{22}{\textbf{\nano\meter}} \fdsoi technology can significantly reduce the energy consumption of small \ao circuits with low-frequency inputs compared to state-of-the-art ultra-low-power CMOS designs. Silicon measurements on an FIR filter and an AES cryptographic core show reduced operating voltage and up to \SI{9.8}{\times} and \SI{1.83}{\times} reductions in leakage power compared to equivalent \hvt and \uhvt CMOS implementations, respectively.
\end{abstract}

\section{Introduction}
\label{sec:intro}

\IEEEPARstart{W}{ith} recent advances in wearable devices and artificial intelligence (AI) algorithms at the edge, there is a growing interest in continuous, real-time sensing applications across multiple domains, ranging from healthcare \cite{health_monitoring, Arrhythmia_predict} to always-on personal assistants with voice activity detection (VAD) \cite{VAD, price2017low}. 
However, the increasing computational demands of AI algorithms required to achieve higher prediction accuracy significantly threaten the autonomy of such systems, which operate under stringent energy and power constraints. Consequently, innovative ultra-low-power design techniques are essential to achieve the required energy efficiency while meeting application timing and accuracy requirements. To address this challenge, edge AI systems are often implemented in \fdsoi technologies \cite{edgeAI-FDSOI-1,edgeAI-FDSOI-2,edgeAI-FDSOI-3, schiavone2021arnold}, which offer an attractive trade-off between performance, power, and energy by reducing leakage and enabling performance–power tuning through body biasing. To leverage the input signal sparsity of real-time applications to further reduce system energy consumption, event-driven architectures \cite{rossi2021vega,wang2021148,liu20214,9233931} have been proposed, in which an \ao domain continuously monitors incoming data and selectively wakes up a high-performance (HP) domain only when a relevant event, such as the detection of a specific pattern or the availability of a new data batch, occurs. The HP domain typically operates near its minimum energy voltage, significantly reducing the energy consumption associated with intensive computation. While this approach effectively reduces energy consumption in the HP domain, it shifts the energy bottleneck to the \ao domain for low-duty-cycle applications operating at kHz input data rates \cite{biomedbench}. In such systems, which spend more than 85\% of their time waiting for events, the leakage power of the \ao domain can dominate the overall energy consumption and directly limits the autonomy of the edge device.

Conventional CMOS logic remains the de facto standard for the design of digital IoT systems due to its mature design methodology, robustness, scalability, and area efficiency. However, CMOS circuits inherently suffer from high leakage currents, which persist even with \hvt or \uhvt transistors in low-power \fdsoi technologies. Although leakage is less critical for compute-intensive workloads, it becomes the dominant power component in ultra-low-power \ao systems running at a reduced frequency, significantly decreasing the battery lifetime of standby \cite{AO_leakage,leakage_reduction_survey}. 
Consequently, critical \ao subsystems, which are often notably smaller than the HP domain, are attractive candidates for alternative logic and design styles that trade area for power efficiency, even if this comes at the expense of additional design effort and complexity.
 
Among such alternatives, \dlsl effectively suppresses subthreshold leakage currents via source biasing, enabling leakage power as low as a few femtowatts per gate in a \SI{180}{\nano\meter} technology node \cite{dlsl_lim20158}. Unfortunately, \dlsl logic exhibits relatively low operating speed due to its feedback-based mechanism, achieving only \SI{6.6}{\hertz} clock frequency in a \SI{32}{}-bit processor implementation \cite{dlsl_lim20158}.  To overcome this limitation, Lin et al. \cite{lin2018dualmode} proposed an enhanced leakage suppression logic family capable of dynamically switching between \dlsl operation and conventional CMOS mode. However, this approach requires additional external control signals and dedicated \ao circuits to generate and manage them, thereby increasing design complexity, area overhead, and leakage power.
Finally, \flsl \cite{cerqueira2018fw,li2021investigation,cerqueira2019femto} was introduced to overcome the speed limitations of \dlsl logic, without requiring additional control signals, but at the cost of increased area. Preliminary studies \cite{li2021investigation} show that FLSL can deliver up to a \SI{28.4}{\times} improvement in operating speed over \dlsl for similar leakage power at the same technology node. 

Although \flsl logic has demonstrated promising results in technology nodes above \SI{90}{\nano\meter} \cite{li2021investigation,cerqueira2019femto}, its effectiveness in sub-\SI{90}{\nano\meter} bulk CMOS technologies remains limited due to the increasing impact of gate and junction leakage currents. In this context, \fdsoi technologies offer a compelling opportunity to unlock the full potential of \flsl logic, as their inherent ability to suppress junction leakage and to enable body-bias performance tuning makes them particularly well-suited for leakage-sensitive ultra-low-power applications. 

In this work, we elaborate on this concept and demonstrate the advantages of \flsl logic in \fdsoi technologies over state-of-the-art ultra-low power CMOS flavors, thereby paving the way for next-generation ultra-low power \ao systems. Unfortunately, existing sizing and characterization methodologies remain immature and lack the accuracy required for a reliable design at advanced nodes. To overcome these limitations, this work also provides a comprehensive design methodology to enable the implementation of functional and efficient \flsl circuits.

Specifically, the main contributions of the presented work on \flsl circuits in \fdsoi technologies are as follows:
\begin{itemize}
  \item we demonstrate that \flsl circuits implemented in \fdsoi technologies restore the leakage suppression capability of \flsl beyond \SI{90}{\nano\meter}, enabling better energy efficiency than state-of-the-art ultra-low-power CMOS technologies;
  \item we show that body biasing further enhances the performance–leakage-power trade-off of \flsl circuits in \fdsoi technologies, providing more efficient tuning capabilities compared to conventional CMOS implementations;
  \item to overcome the lack of systematic sizing and characterization methodologies for \flsl, we propose a comprehensive characterization and step-by-step transistor-sizing methodology that enables designers to achieve optimal performance in the \flsl logic family;
  \item we present an adapted digital design flow that enables the conception of functionally working always-on \flsl circuits on silicon;
  \item we validate the proposed approach with silicon fabrication and measurements in complete \flsl circuits, showing that \flsl achieves a leakage reduction of up to 1.83$\times$ compared to the state-of-the-art \uhvt CMOS technology with \rbb.
\end{itemize}

The remainder of this paper is organized as follows. 
\cref{sec:background} reviews related work on CMOS and leakage suppression logic techniques for ultra-low-power \ao circuits. 
\cref{sec:fdsoi} analyzes the impact of \fdsoi technology on the leakage reduction mechanisms of \flsl and its capability to enable leakage-performance trade-off through body biasing. 
\cref{sec:design} describes the characterization, transistor sizing, and implementation methodology used to design ultra-low-power \flsl circuits.
\cref{sec:silicon} presents silicon measurement results of \flsl circuits implemented in a 22 nm \fdsoi process, validating the benefits of \fdsoi-based \flsl circuits compared to ultra-low-power CMOS implementations.
Finally, \cref{sec:conclusion} summarizes the main conclusions from this work.

\section{Related works}
\label{sec:background}

\fdsoi technologies provide a favorable energy–performance trade-off for the conception of ultra-low-power \ao circuits that are dominated by leakage power by suppressing junction leakage currents and enabling performance–power tuning through body biasing. Orthogonally, to further reduce energy consumption, recent work has also explored different CMOS design techniques and alternative logic styles that can be applied in \fdsoi technologies. 

\subsection{Ultra-low-power Always-On CMOS design techniques}

To mitigate leakage currents in always-on CMOS circuits, designers commonly adopt standard cells based on \hvt and \uhvt transistors. Indeed, higher threshold voltages drastically reduce subthreshold currents due to their exponential dependence on the threshold voltage. Such approaches are particularly well suited for leakage-sensitive always-on components such as wake-up controllers \cite{wakeup_HVT}, near-sensor processing circuitry \cite{MFCC_HVT}, and custom SRAMs for data retention \cite{AO-SRAM-UHVT}. Using foundry-verified design kits, this approach enables fast and reliable implementation while maintaining strong compatibility with the standard design flow used for the rest of the system.  
In \fdsoi, \rbb can also be applied to further increase the effective threshold voltage, but the body bias voltage must be managed by a dedicated on-chip or external voltage regulator, leading to higher design complexity, area, and leakage power, which may outweigh the expected energy savings for small \ao circuits.
Despite the effectiveness of this method, increasing the threshold voltage of transistor devices is also fundamentally limited by the value of the supply voltage required to ensure reliable operation. This limitation is further exacerbated in advanced technology nodes, where the supply voltage is reduced according to Dennard's scaling law \cite{dennard}, and therefore limits the upper limit of the threshold voltage for devices. As a result, leakage currents in CMOS devices remain high and are ultimately limited by subthreshold conduction, as discussed in \cref{sec:fdsoi}.

To effectively reduce leakage currents in CMOS circuits, aggressive voltage scaling in the subthreshold regime can also be explored \cite{lotze201162,hwang200785mv,NVTH-lib} for low-frequency standby operations. This approach has been proven to be one of the most efficient techniques for CMOS-based circuits where processing speed is not a stringent requirement. Indeed, the \ac{emv} for CMOS-circuits usually sits in the subthreshold region where maximum energy savings can be achieved\cite{model-subVt}. However, circuit operation in the subthreshold regime requires careful design considerations to ensure proper functionality \cite{NVTH-lib}, as the exponential decrease in transistor drive current leads to large variations in gate delay. This makes it extremely challenging to meet real-time timing constraints, particularly for control and wake-up logic that must respond within bounded latency. In addition, operations in the subthreshold regime increase sensitivity to process, voltage, and temperature (PVT) variations, which can compromise design robustness and increase susceptibility to environmental fluctuations. Finally, while dynamic power scales quadratically with the supply voltage, leakage power does not, limiting achievable energy savings for leakage-dominated always-on circuits. 

\subsection{Leakage Suppression Logic}

\begin{figure}[!t]  
    \centering
    \begin{subfigure}[b]{0.48\columnwidth}
        \centering
        \includegraphics[width=\linewidth]{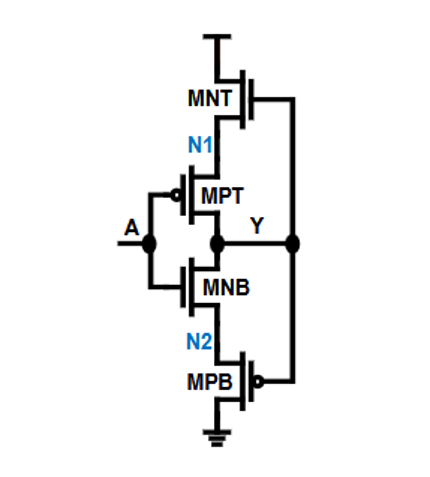}
        \caption{\dlsl inverter \cite{dlsl_lim20158}}
        \label{fig:dlsl_inv}
    \end{subfigure}
    \hfill
    \begin{subfigure}[b]{0.48\columnwidth}
        \centering
        \includegraphics[width=\linewidth]{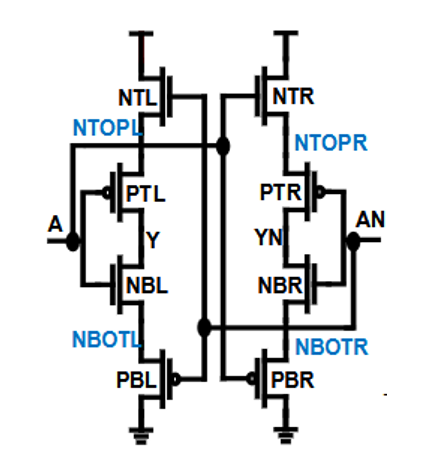}
        \caption{\flsl inverter \cite{cerqueira2018fw}}
        \label{fig:flsl_inv_soa}
    \end{subfigure}
     \hfill
    \begin{subfigure}[b]{0.48\columnwidth}
        \centering
        \includegraphics[width=\linewidth]{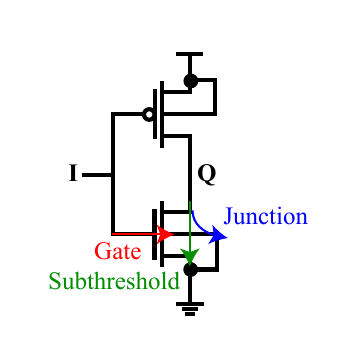}
        \caption{CMOS inverter}
        \label{fig:Leak_definition}
    \end{subfigure}
    \caption{Schematics of a \dlsl (a), \flsl (b), and CMOS inverter (c).}
    \label{fig:BG_inverter}
\end{figure}

In response to CMOS-based \ao circuit limitations, Lim et al. \cite{dlsl_lim20158} introduced a new logic family known as \dlsl. A schematic of a \dlsl inverter is shown in \cref{fig:dlsl_inv}. Additional NMOS headers and PMOS footers are added to the conventional CMOS gate, with their gate terminals connected to the output node. This feedback mechanism enables the additional transistors to enter a super-cutoff state, thereby significantly reducing subthreshold leakage. However, this topology suffers from extremely low operating speed (on the order of Hz) due to the feedback path, which severely limits the set of applications it supports.

To overcome this limitation, Lin et al. \cite{lin2018dualmode} proposed a novel logic family that combines CMOS and \dlsl logic for battery-indifferent distributed sensing. This logic family can dynamically switch between \dlsl operation and conventional CMOS mode based on the energy harvested by a solar panel. Although the implemented microcontroller achieves clock frequencies of a few MHz in CMOS mode, operation in the \dlsl mode is still limited to only a few hertz. As a result, circuit performance becomes highly dependent on external environmental conditions, posing a challenge for real-time applications subject to stringent timing constraints. Moreover, this implementation requires additional control logic and power-management circuitry to generate optimal supply voltages for each operating mode. For small always-on circuits, such overhead can outweigh the achieved energy savings, making the approach inefficient.

To overcome the very low switching speed of the \dlsl logic and the use of extra control logic, Cerqueira et al. introduced the \flsl family \cite{cerqueira2018fw,cerqueira2019femto}. \cref{fig:flsl_inv_soa} shows the schematic of an \flsl inverter where the slow feedback path used in \dlsl to activate the double source biasing from the output node has been replaced by a dual-rail implementation. In this way, the authors achieve significant improvements in both operating speed and leakage-power delay product (LDP) compared to \dlsl, with gains of up to \SI{148}{\times} and \SI{1982}{\times}, respectively, in a \SI{180}{\nano\meter} technology node at the cost of a \SI{33}{\percent} area overhead. Demonstrated with an always-on 16-bit, 8-tap finite impulse response (FIR) filter, the authors achieved a leakage power as low as \SI{109}{\pico\watt} with a maximum operating frequency of \SI{1.03}{\kilo\hertz}. 
 
Building on these promising results, subsequent work has explored the impact of technology scaling on \flsl logic, aiming to further improve its operating speed and expand its applicability to a broader range of ultra-low-power edge applications, such as neural signal processing, filtering, and keyword spotting. Most notably, Li et al. \cite{li2021investigation} investigate and compare the performance of \dlsl and \flsl across nine technology nodes, ranging from \SI{180}{\nano\meter} bulk CMOS to \SI{7}{\nano\meter} FinFET. Their study highlights that \flsl offers an interesting trade-off: technology nodes above \SI{90}{\nano\meter} benefit from higher leakage power reduction by efficiently reducing subthreshold leakage currents compared to CMOS logic, while more advanced technology nodes enable significant improvements in operating speed up to \SI{2} to \SI{3} orders of magnitude higher at the cost of less effective leakage suppression capabilities due to the dominance of gate and junction leakage currents in advanced technology nodes. Interestingly, prior work has been limited to bulk and FinFET technology nodes and has not assessed \flsl performance in \fdsoi technologies. This paper overcomes this limitation by demonstrating the suitability of \fdsoi for \flsl. We also provide a detailed analysis of the \flsl operating mechanisms and a systematic sizing strategy for reaching optimal performance.

\section{Leveraging \fdsoi for \flsl}
\label{sec:fdsoi}

With continued transistor scaling, conventional CMOS always-on circuits become increasingly inefficient due to increasing leakage currents, dominated by subthreshold and junction components. While \fdsoi technologies effectively suppress junction leakage, \flsl can be leveraged to drastically reduce subthreshold leakage. This section demonstrates the benefits of combining \flsl logic with \fdsoi technologies, despite the inefficiencies previously reported for \flsl in sub-\SI{90}{\nano\meter} nodes.

\subsection{Enhancing Leakage Suppression}
Due to their well-established design methodologies, low fabrication cost, high robustness, and area efficiency, bulk CMOS technologies have long been the default choice for implementing \ao circuits. However, their inherently high leakage currents hinder designers from achieving the nanowatt power levels that are required by ultra-low-power edge devices, which significantly limits system autonomy. 

\begin{figure}[htb]
    \centering
        \includegraphics[width=\linewidth]{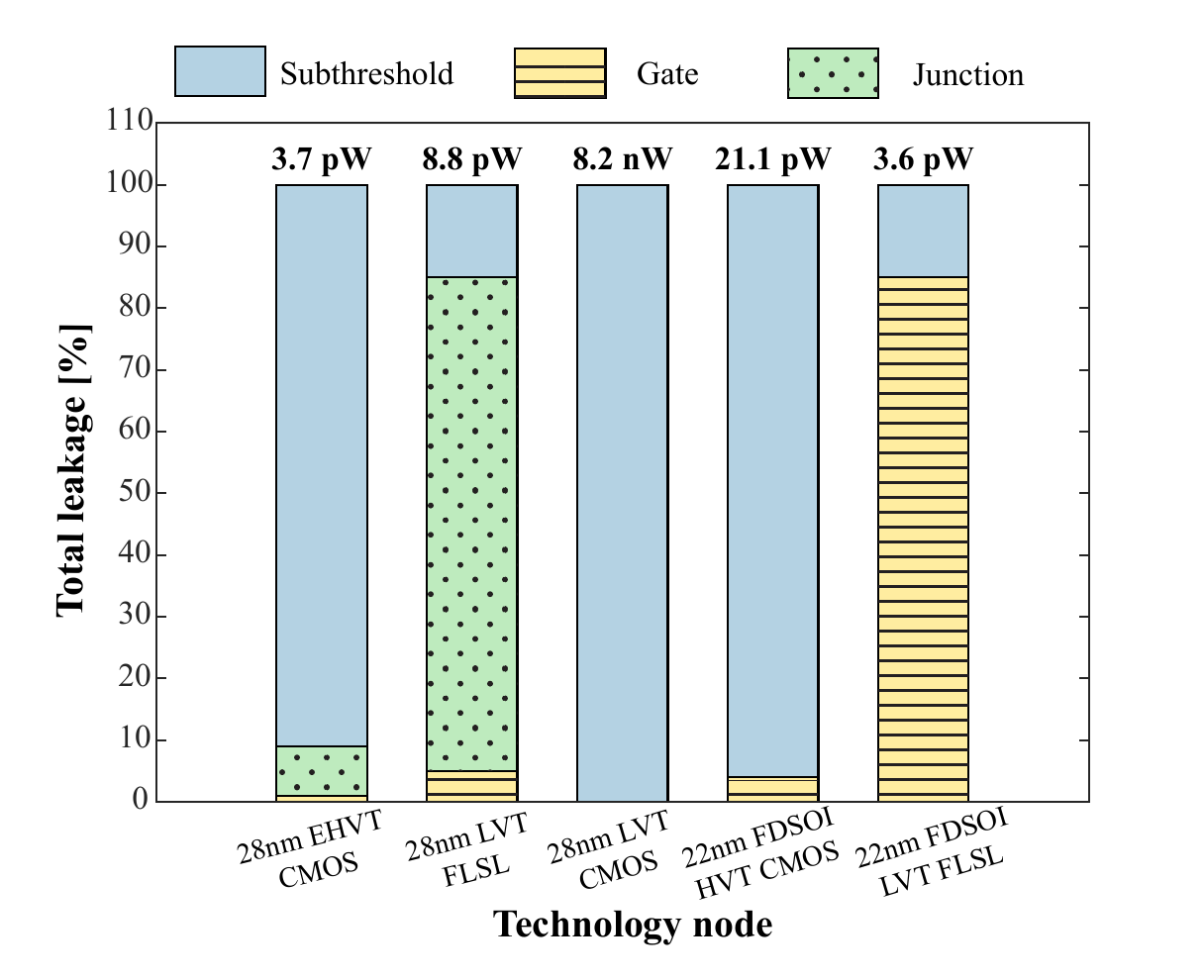}
        \caption{Leakage contributions at nominal voltage for CMOS and \flsl logic on a 28nm bulk and a 22nm \fdsoi technology node.}
        \label{fig:Leak_contr}
\end{figure}

As illustrated in \cref{fig:Leak_definition}, the leakage current in CMOS logic can be divided into three primary components:
\begin{itemize}
\item \textit{Subthreshold leakage} between the drain and source of a transistor; this is typically the dominant contributor to total leakage.
\item \textit{Junction leakage} arising from the PN junctions between the source/drain regions and the substrate.
\item \textit{Gate leakage} from tunneling through the gate oxide.\\
\end{itemize}
To solve the high leakage of CMOS logic, Cerqueira et al. introduced the \flsl logic style to mitigate subthreshold leakage by exploiting source bias. \flsl is usually implemented using \lvt transistors to achieve higher operating speed, allowing a wider range of applications \cite{cerqueira2019femto}.
The impact of \flsl on the different leakage mechanisms is illustrated in \cref{fig:Leak_contr}, which compares the leakage contributions of different CMOS and \flsl inverter implementations in a \SI{28}{\nano\meter} bulk and a \SI{22}{\nano\meter} \fdsoi technology.
Implemented with \lvt and \ehvt transistors, the leakage power of the CMOS inverters ranges from \SI{3.7}{\pico\watt} to \SI{8.2}{\nano\watt} and is largely dominated by subthreshold leakage, which represents up to 99.96\% of total leakage in the \lvt configuration. In contrast, the \flsl inverter implemented in the same technology using \lvt transistors drastically reduces the subthreshold leakage contribution to 15\% and achieves a total leakage power of \SI{8.8}{\pico\watt}, corresponding to an improvement of 930× compared to its CMOS counterpart. However, this value still remains \SI{2.4}{\times} higher than that of a CMOS inverter implemented with \ehvt transistors, indicating that \flsl is not a competitive choice in this technology. 

This limitation is due to high junction leakage, which accounts for up to 80\% of the total \flsl leakage. This leakage is mainly due to the use of \lvt transistors and technology scaling, which reduce both the thickness of the gate oxide and the length of the channel, thus promoting gate oxide tunneling and enhancing junction leakage through increased carrier tunneling. These effects ultimately limit the leakage-reduction capability of \flsl in sub-\SI{90}{\nano\meter} nodes, as reported by Li et al. \cite{li2021investigation}.

However, \fdsoi technologies incorporate a thin buried oxide layer that suppresses the parasitic PN junction with the substrate and prevents current leakage from the channel to the substrate. Consequently, junction leakage is already effectively suppressed, significantly improving the leakage reduction capacity of \flsl. This behavior is illustrated in the last two columns of \cref{fig:Leak_contr}, which compares the leakage contributions of an \flsl-\lvt inverter and our \hvt-CMOS inverter implemented in a \SI{22}{\nano\meter} \fdsoi node. 
This time, the \flsl inverter achieves a total leakage power of \SI{3.6}{\pico\watt}, corresponding to an improvement by a factor of \SI{5.8}{\times} compared to the \hvt-CMOS inverter. Consequently, \flsl remains competitive and even superior in terms of leakage compared to conventional CMOS logic in \fdsoi technology. These results indicate that \flsl is a promising alternative to conventional CMOS logic for ultra-low-power \ao circuits in \fdsoi technologies and motivate a deeper analysis and understanding of the \flsl mechanism to fully exploit its potential. 

\subsection{Performance-Leakage trade-off with body biasing}

\begin{figure*}[tb]
    \begin{subfigure}{.33\textwidth}
        \centering
        \includegraphics[width=\linewidth]{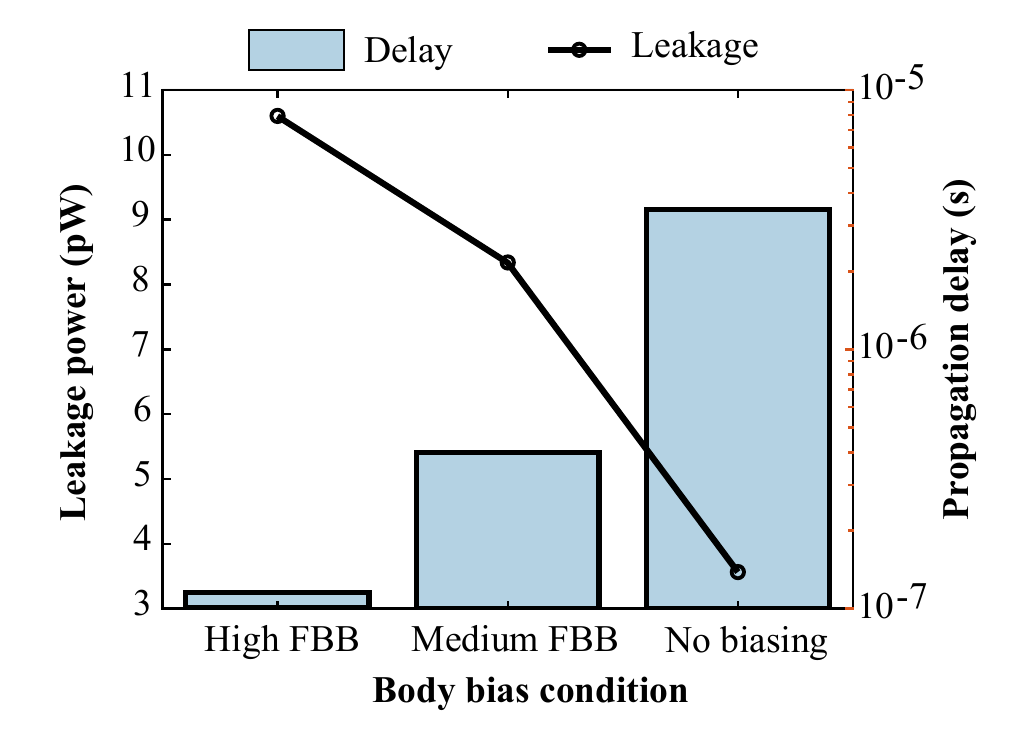}
         \caption{\flsl inverter}
         \label{fig:body-biasing-tradeoff}
    \end{subfigure}\hfill
    \begin{subfigure}{.33\textwidth}
        \centering
        \includegraphics[width=\linewidth]{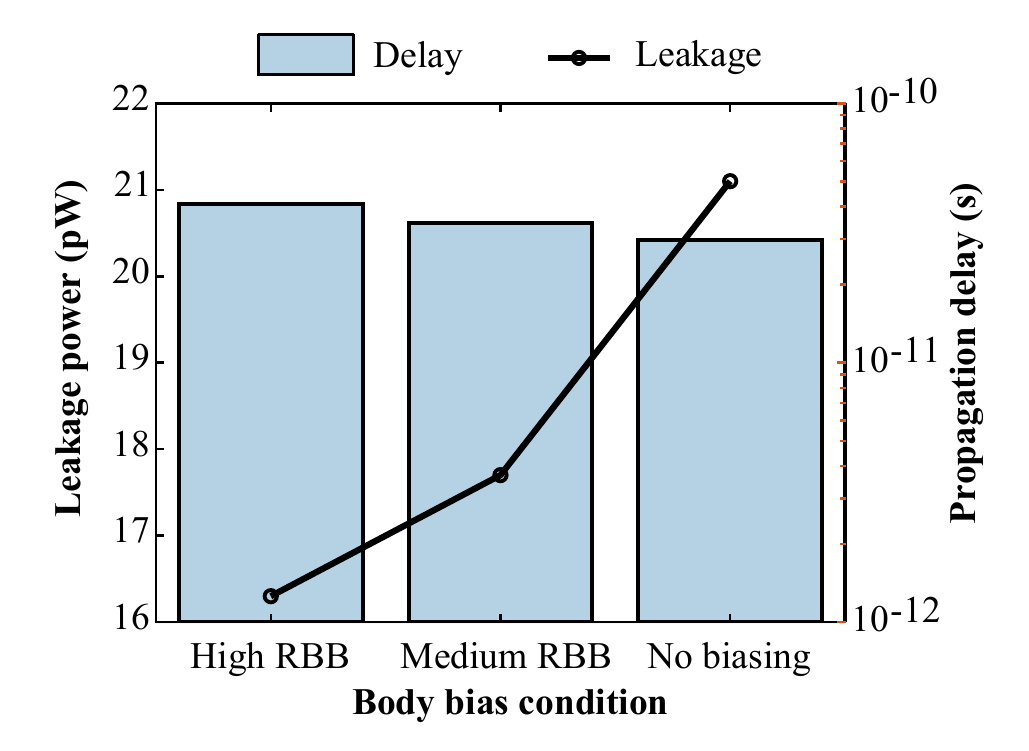}
        \caption{\hvt CMOS inverter}
        \label{fig:body-biasing-tradeoff-CMOS}
    \end{subfigure}\hfill
    \begin{subfigure}{.305\textwidth}
        \centering
        \includegraphics[width=\linewidth]{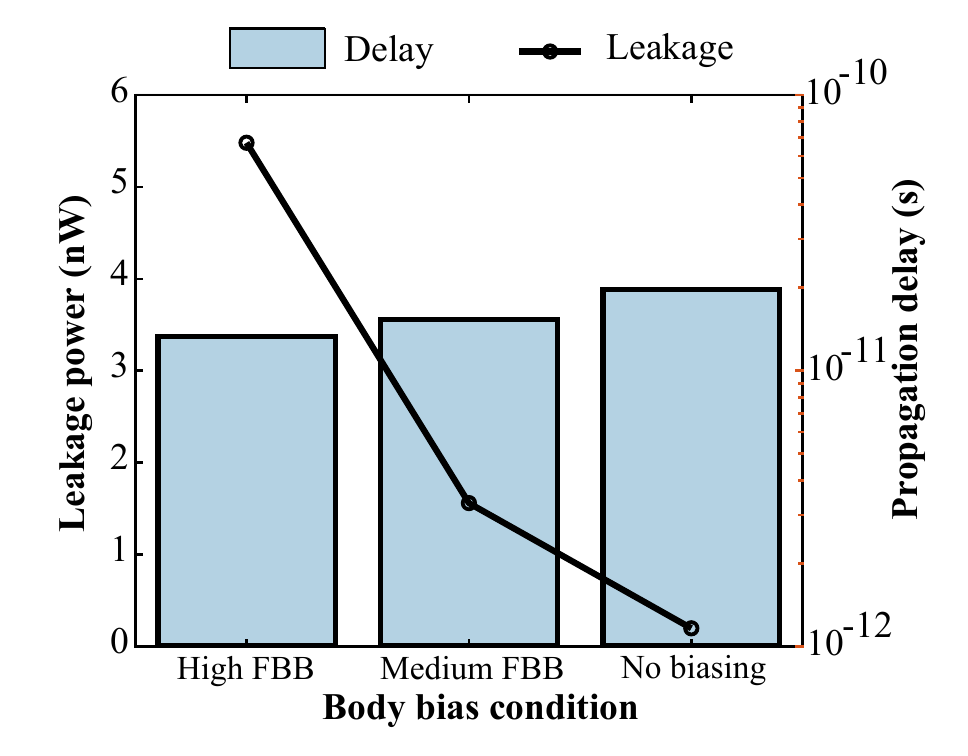}
        \caption{\lvt CMOS inverter}
        \label{fig:body-biasing-tradeoff-CMOS-LVT}
    \end{subfigure}
    \caption{FO4 delay (bars) and leakage power (line) trade-offs for an \flsl inverter (a), an \hvt CMOS inverter (b) and a \lvt CMOS inverter (c) under body biasing on \fdsoi substrate.}
    \label{fig:bb-perf-cmos}
    \vspace{-2mm}
\end{figure*}

In the previous section, we demonstrated that \flsl can serve as a promising alternative to conventional CMOS logic in \fdsoi technologies to reduce leakage power in low-frequency \ao circuits. Nonetheless, a key advantage of \fdsoi over bulk technologies is the ability to tune performance depending on application requirements thanks to body biasing. Therefore, body biasing must be investigated for both CMOS and \flsl to assess the respective impacts.

For \flsl gates implemented with \lvt devices, \fbb is applied by biasing the NMOS and PMOS bodies with positive and negative voltages, respectively. The resulting performance trade-offs for an \flsl inverter are shown in \cref{fig:body-biasing-tradeoff}. Applying a strong \fbb improves the switching speed of the \flsl inverter by more than \SI{30}{\times}, while it increases its total leakage by only \SI{3}{\times} compared to the unbiased case. This demonstrates that significant performance gains can be achieved at relatively low leakage cost and at an overall \SI{10}{\times} energy advantage.

A similar analysis is performed for CMOS inverters implemented with \lvt or \hvt transistors, as shown in \cref{fig:body-biasing-tradeoff-CMOS} and \cref{fig:body-biasing-tradeoff-CMOS-LVT}. For the \lvt-CMOS inverter, strong \fbb leads to significantly worse overall energy efficiency, with a \SI{27}{\times} increase in leakage for only a \SI{1.5}{\times} delay improvement. Therefore, the \flsl inverter demonstrates superior performance tuning compared to its CMOS counterpart. 

This improvement comes from the use of \lvt devices in the \flsl logic, which significantly reduces propagation delay under \fbb while introducing only a limited leakage overhead. Indeed, \lvt transistors significantly affect the \flsl switching mechanism, which relies on both fast and slow transition regimes, as demonstrated later in \cref{delay-characterization}. 
In the fast transition regime, \lvt devices enable functional transistors to operate more deeply in the linear region by providing a higher gate overdrive, thereby accelerating switching. In the slow transition regime, they increase the leakage currents through the header and footer transistors, which speeds up the discharge of internal nodes. Together, these effects result in a substantial improvement in switching speed.
Despite this significant performance gain, the increase in leakage current remains well controlled thanks to the efficient leakage reduction provided by source biasing, since this mechanism is strongly dependent on the supply voltage, which determines the negative gate overdrive applied to the transistors. In contrast to the \flsl inverter, the \hvt-CMOS inverter shows limited sensitivity to body bias: applying \rbb yields only a \SI{1.3}{\times} leakage reduction at the cost of a \SI{1.4}{\times} delay penalty, resulting in nearly proportional trade-offs. In general, these results demonstrate that \flsl offers significantly more efficient body-bias-driven tuning than conventional CMOS, allowing for large performance gains at a modest leakage cost as a highly effective knob for optimizing the power-delay product.

\section{\flsl design method}
\label{sec:design}

\begin{figure*}
    \centering
        \includegraphics[width=1\linewidth]{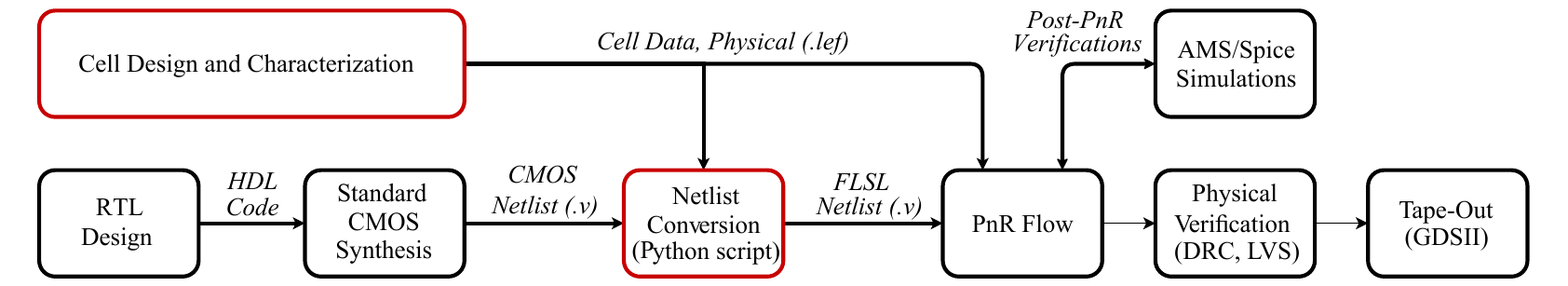}
        \caption{Block diagram of the adapted digital design flow used for \flsl circuit implementation. Steps circled in red differ from conventional CMOS digital flow}
        \label{fig:flsl_flow}
\end{figure*}
Following the demonstration in the previous section of the benefits of \fdsoi for \flsl logic, a deeper understanding of the \flsl operating mechanism is required to fully exploit its potential. However, previous studies lack such a comprehensive analysis, and because of the differential logic style, conventional CMOS propagation delay models and design methodologies cannot be applied directly to \flsl circuits. Therefore, we first characterize and design the basic \flsl cells, providing a thorough understanding of their switching mechanism. Based on this characterization, we define a methodology that guides designers towards optimal gate sizing to minimize the \ldp, a key metric for always-on circuits where leakage current is the dominant source of power consumption. Once the basic \flsl library is established, it can be integrated into a conventional synthesis and flow through the route and place with minimal custom adjustments, as illustrated in \cref{fig:flsl_flow}.

\subsection{\flsl switching mechanism and delay}
 \label{delay-characterization}
Since the \flsl logic family relies on subthreshold conduction and a differential input/output structure, the CMOS propagation delay, which is typically defined as the time interval between the \SI{50}{\percent} voltage crossing of the input signal and the corresponding \SI{50}{\percent} crossing at the output, does not adequately capture the \flsl switching behavior.
As a result, an adapted delay characterisation approach is required to assess the propagation delay of \flsl circuits.
 
By construction, any \flsl gate comprises a "direct path" with non-inverted inputs and outputs (e.g., $\mathrm{I}$ and $\mathrm{Q}$) and an "inverted path" with inverted inputs and outputs (e.g., $\bar{\mathrm{I}}$ and $\bar{\mathrm{Q}}$). Additional transistors are added as headers for the CMOS pull-up network and as footers for the CMOS pull-down network. 
We divide the transistors into two categories:  
\begin{itemize}
    \item Functional transistors: active transistors responsible for the correct functionality of the logic gate and part of the conventional CMOS pull-up or pull-down network.
    \item Header/Footer transistors: additional NMOS (header) and PMOS (footer) transistors mainly operating in their respective super-cutoff regimes to limit subthreshold leakage current. \\
\end{itemize}

\begin{figure}
    \centering
        \includegraphics[width=0.9\linewidth]{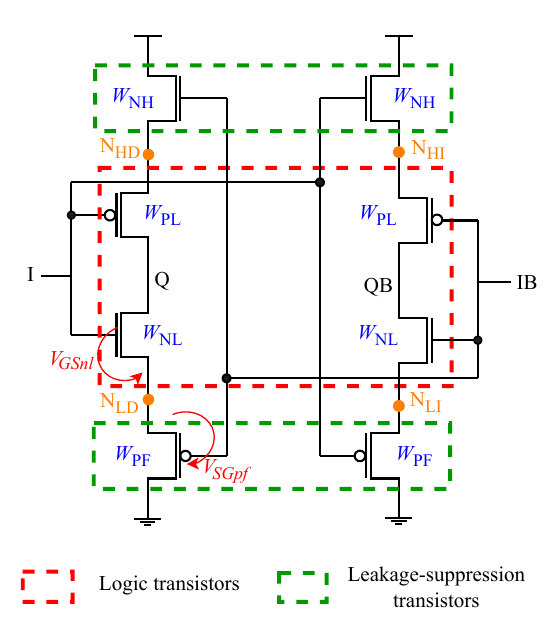}
        \caption{Schematic of an \flsl inverter annotated with the conventions used in this work.}
        \label{fig:flsl_inv}
\end{figure}

For illustration, a schematic of an \flsl inverter is shown in \cref{fig:flsl_inv}. In the following, we use this inverter as a reference to establish our characterisation and design methodology. However, the proposed methodology can be extended to any \flsl gate through the following definitions that apply to all single-stage negative-unate boolean gates:

\begin{itemize}
    \item $W_{\mathrm{NL}}$ and $W_{\mathrm{PL}}$ denote the widths of all functional NMOS and PMOS transistors.
    \item $W_{\mathrm{NH}}$ and $W_{\mathrm{PF}}$ denote the widths of all NMOS header and PMOS footer transistors.
    \item $\mathrm{N_{LD}}$ and $\mathrm{N_{LI}}$ refer to the "low" intermediate nodes located between the pull-down network and the first PMOS footer of the direct and inverted path, respectively.
    \item  $\mathrm{N_{HD}}$ and $\mathrm{N_{HI}}$ refer to the "high" intermediate nodes located between the pull-up network and the first NMOS header of the direct and inverted path, respectively.\\
\end{itemize}

\begin{figure}
    \centering
     \begin{subfigure}{.4\textwidth}
        \centering
        \includegraphics[width=\linewidth]{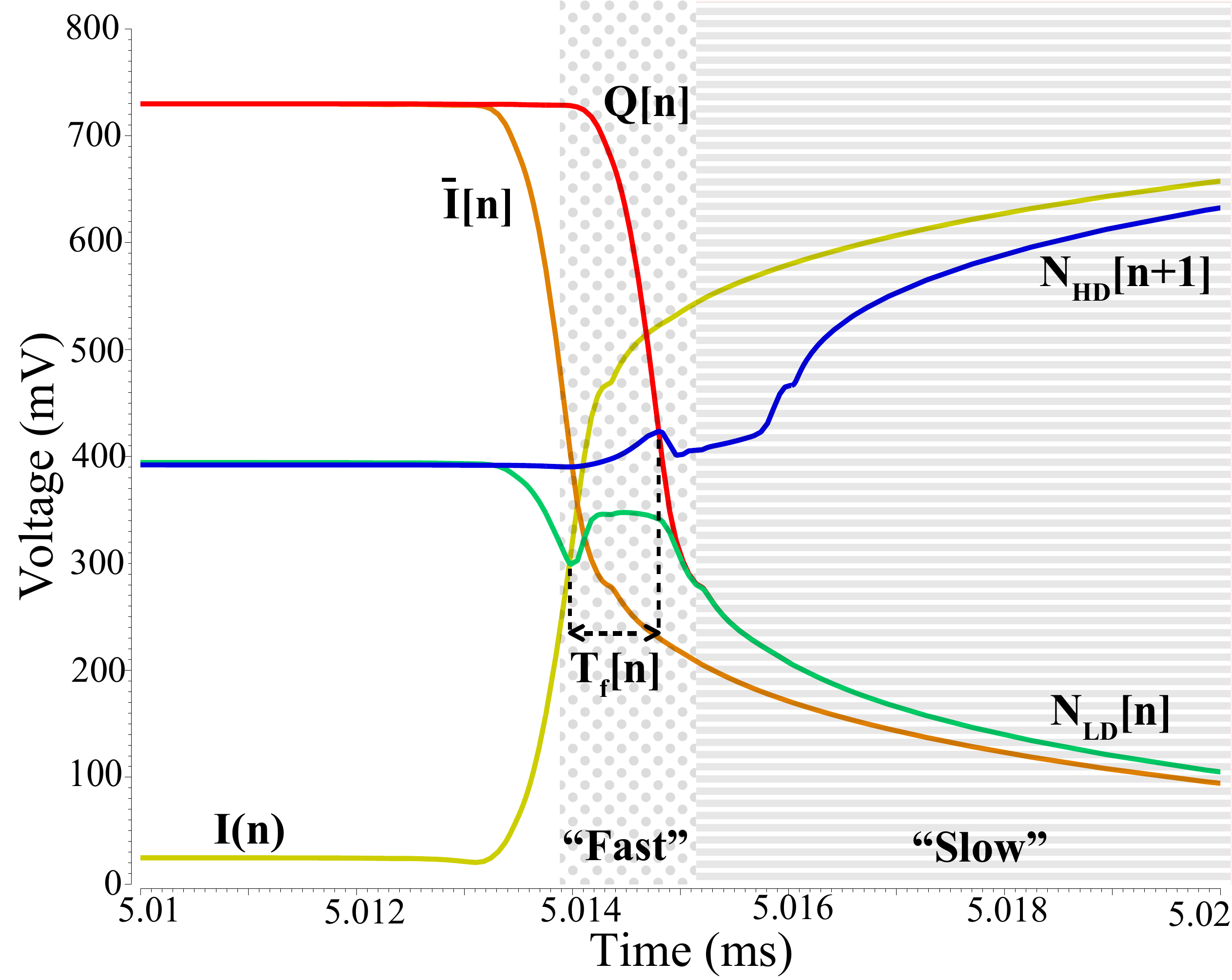}
        \caption{}
        \label{fig:I_rise}
    \end{subfigure}
    \vspace{0.5cm}  
    \begin{subfigure}{.4\textwidth}
        \centering
        \includegraphics[width=\linewidth]{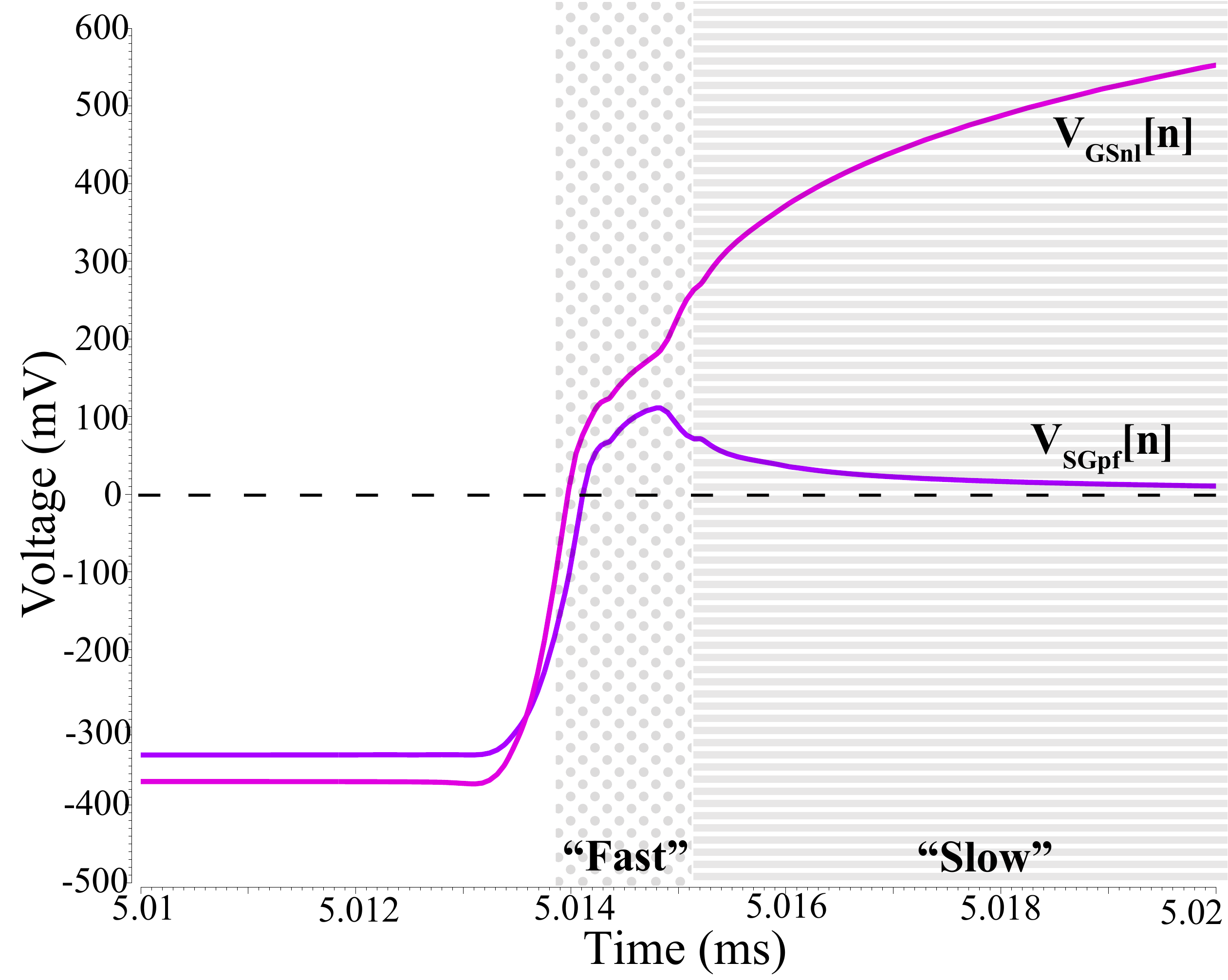}
        \caption{}
        \label{fig:I_fall}
    \end{subfigure}
    \caption{ \flsl waveforms corresponding to a falling output edge.}
    \label{fig:flsl_waveforms}
\end{figure}

For the analysis of the \flsl switching mechanism, we consider an \flsl inverter $n$ that drives another \flsl inverter $n+1$, and we analyse the case corresponding to a rising input $\mathrm{I}[n]$ to gate $n$, but similar reasoning can be applied for a falling input as well. 
Examples of the corresponding relevant waveforms are shown in \cref{fig:I_rise} and \cref{fig:I_fall}. 
As reported by \cite{cerqueira2018fw}, the decrease of the corresponding inverted input $\bar{\mathrm{I}}[n]$ rapidly raises the gate voltage $\mathrm{V_{GSpf}}[n]$ of the PMOS footer transistor, which transitions the latter from a super-cutoff state to a cutoff state. This change allows $\mathrm{N_{LD}}[n]$ to discharge through the PMOS footer, helping the NMOS functional transistor to transition from a super-cutoff state to a cutoff state. Then, the fast transition on the direct output node $\mathrm{Q}[n]$ (dotted area highlighted in \cref{fig:flsl_waveforms}) occurs when both the gate voltage of the PMOS footer $\mathrm{V_{GSpf}}[n]$ and the gate voltage of the functional NMOS $\mathrm{V_{GSnl}}[n]$ are greater than \SI{0}{\volt}. The latter causes the direct output $\mathrm{Q}[n]$ to discharge through the functional NMOS transistor until it reaches the same voltage as the intermediate node $\mathrm{N_{LD}}[n]$. Finally, both $\mathrm{Q}[n]$ and $\mathrm{N_{LD}}[n]$ nodes slowly discharge towards ground due to subthreshold leakage of the footer transistors.
From this example, we see that an \flsl transition can be characterized by the succession of \SI{2}{} different regimes: 
\begin{itemize}
    \item a "fast transition" regime induced by the functional transistors which progressively turn on as soon as the input voltage reaches the voltage of "low" or "high" intermediate nodes (i.e., $\mathrm{N_{LD}}[n]$ or $\mathrm{N_{HD}}[n]$).
    \item a "slow transition" regime induced by the leakage currents of footer and header transistors.
\end{itemize}

Considering that the propagation delay through a gate $n$ is best defined as the time between a change in input voltage activating the output transition and the corresponding change in output voltage required to activate the gate $n+1$, the fall time $T_\mathrm{f}[n]$ and rise time $T_\mathrm{r}[n]$ for an \flsl gate $n$ can be defined as follow when assuming a rising edge on $\mathrm{I}[n]$:
\begin{equation}
\label{eq:tr}
    \small
    T_\mathrm{f}[n] = T(\mathrm{Q}[n]=\mathrm{N_{HD}}[n+1]) - T(\mathrm{I}[n]=\mathrm{N_{LD}}[n])
\end{equation}
\begin{equation}
\label{eq:tf}
    \small
    T_\mathrm{r}[n] = T(\bar{\mathrm{Q}}[n]=\mathrm{N_{LI}}[n+1]) - T(\bar{\mathrm{I}}[n]=\mathrm{N_{HI}}[n]),
\end{equation}
where $T(\mathrm{A}=\mathrm{B})$ corresponds to the time when the voltage of node $\mathrm{A}$ is equal to the voltage of node $\mathrm{B}$.\\

Due to the differential structure of \flsl gates, signal propagation is characterized by both a rising and a falling transition on the direct and on the inverted path. However, the \flsl propagation delay is not equal to the slowest delay between the fall time $T_\mathrm{f}[n]$ and the rise time $T_\mathrm{r}[n]$. Indeed, at the circuit level, the earlier transition already initiates the activation of the subsequent gate by altering the state of the header or footer transistor, pushing it out of super-cutoff mode. This, in turn, triggers the charge or discharge of intermediate nodes, resulting in a faster activation of the next stage.
Empirically, we find that the propagation delay $T_{\mathrm{delay}}[n]$ of an \flsl gate $n$ is well represented by the average of the rise and fall times 
\begin{equation}
    \small
    T_{\mathrm{delay}}[n] = (T_\mathrm{r}[n] + T_\mathrm{f}[n]) / 2
\end{equation}

\subsection{\flsl transistor sizing}
\label{method}

As discussed earlier, both functional transistors and header/footer transistors influence the propagation delay of an \flsl gate, although they affect it in different ways. Consequently, these transistors must be sized independently and carefully, with respect to each other, to optimize the overall performance of the gate. The objective is to optimize the \ldp of the gate.
To achieve this objective, we propose a three-step process in order to achieve optimal transistor sizing. 
\\
\subsubsection{Optimal voltage supply $V_{\mathrm{opt}}$}
\label{step1_method}
Previous studies \cite{cerqueira2018fw,cerqueira2019femto} on \flsl gates have shown that \ldp can be strongly affected by the supply voltage by up to three orders of magnitude.
Therefore, careful selection of the supply voltage is essential to meet the required performance under application-specific constraints.
The significant dependence of the \ldp on supply voltage is primarily attributed to the gate static leakage, as opposed to the propagation delay, which for \flsl remains within the same order of magnitude across a reasonable supply voltage range and is mainly dictated by the technology node. Indeed, the \flsl leakage reduction mechanism highly depends on the strength of the super-cutoff state of header and footer transistors. The latter varies in a non-linear manner with respect to the amplitude of the gate-to-source voltage and, therefore, directly depends on the supply voltage value. In contrast, functional transistors, which are mainly responsible for the fast output transition as highlighted earlier, operate in the subthreshold and in the linear region by cell construction, resulting in a weaker dependence on voltage scaling.
As a result, determining the optimal supply voltage, denoted as $V_{\mathrm{opt}}$, needs to be done before any transistor sizing to minimize the \ldp by achieving the lowest leakage power. The optimal voltage $V_{\mathrm{opt}}$ is identified by performing a DC voltage sweep on a \SI{50}{}-stage \flsl inverter chain. The number of stages has been chosen experimentally to extract with precision leakage current in the range of \SI{}{\femto\watt}.

\begin{figure}[h]
    \centering
        \includegraphics[width=0.9\linewidth]{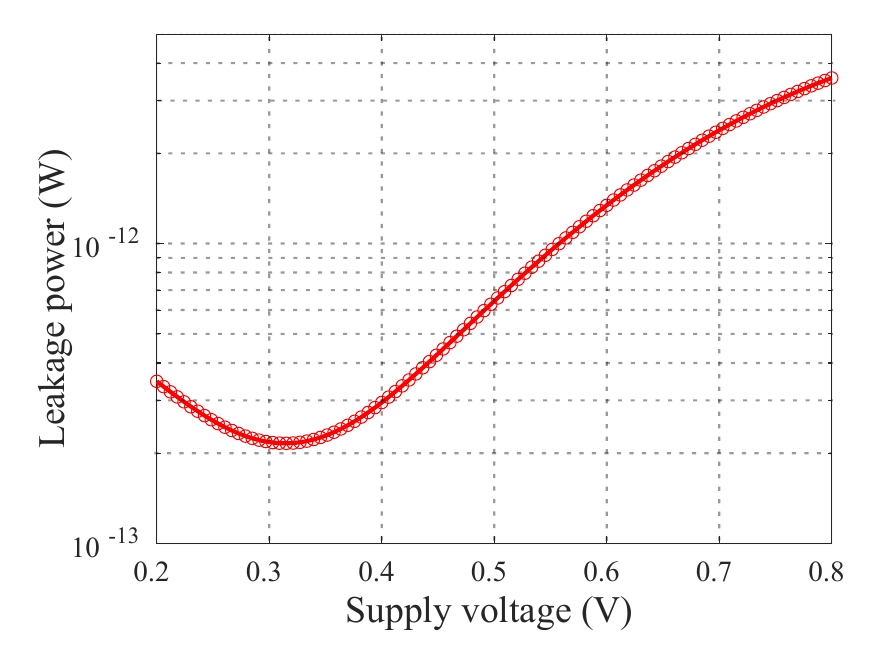}
        \caption{Leakage power versus supply voltage for an \flsl inverter made of \lvt transistors in \SI{22}{\nano\meter} \fdsoi process without body biasing.}
        \label{fig:Vleak}
\end{figure}

\cref{fig:Vleak} shows the leakage power variation versus supply voltage for an \flsl inverter built with \lvt transistors in a \SI{22}{\nano\meter}-\fdsoi process. An optimum voltage $V_{\mathrm{opt}}$ can be identified around \SI{0.3}{\volt} to minimize the gate leakage power. On one hand, for voltages smaller than $V_{\mathrm{opt}}$, the leakage power, dominated by subthreshold leakage currents, increases due to less effective source biasing with reduced negative gate voltage in the header and footer transistors. On the other hand, for supply voltages higher than $V_{\mathrm{opt}}$, leakage currents increases rapidly due to the gate oxide leakage currents. As the supply voltage rises further, this component progressively dominates the subthreshold leakage of the inverter as suggested previously by \cref{fig:Leak_contr}, thereby reducing the effectiveness of \flsl source biasing.\\

\subsubsection{Functional transistor sizing}
\label{step2_method}
\begin{figure}[h]
    \centering
        \includegraphics[width=0.9\linewidth]{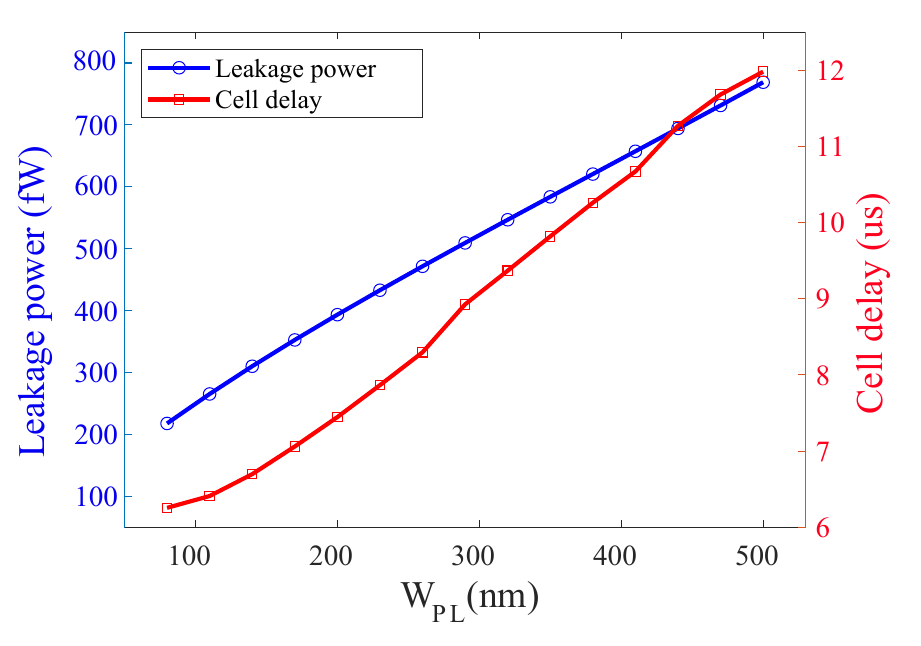}
        \caption{Leakage power and cell delay variations against PMOS functional width $W_{\mathrm{PL}}$ for a \fdsoi \flsl inverter.}
        \label{fig:Wp_study}
\end{figure}

After determining $V_{\mathrm{opt}}$, the next step is to determine the optimal widths for each transistor of the \flsl gate. As shown by \cref{fig:I_rise} and \cref{fig:I_fall}, the gate propagation delay is primarily dominated by the fast transition regime driven by the functional transistors, which are responsible for the movement of charges between the intermediate nodes and the output node. Therefore, these transistors must be sized first. Since functional transistors share their drain terminals with the gate output node and mostly operate in a linear regime, wider transistors should result in a linear degradation of the \ldp with the growing transistor width due to higher leakage current and delay induced by a faster growth of the output capacitance compared to the drive strength. To verify this assumption, \cref{fig:Wp_study} reports leakage and delay variations with respect to the width $W_{\mathrm{PL}}$ of the functional PMOS obtained at $V_{\mathrm{opt}}=0.3V$ for an \flsl inverter built with \lvt transistors. As expected, both leakage power and cell delay increase linearly with the upsizing of $W_{\mathrm{PL}}$ when other transistors are minimum size. 
As a result, both functional transistor widths $W_{\mathrm{NL}}$ and $W_{\mathrm{PL}}$ need to be kept at their minimum sizes ($W_{\mathrm{min}}$), contrary to conventional CMOS sizing, which balances the difference in carrier mobility between NMOS and PMOS transistors with PMOS width upsizing.\\

\subsubsection{Header/Footer transistors sizing}
\label{step3_method}
The final step determines the optimal sizing of the header and footer transistors. The size of these transistors can influence the leakage and cell delay of the gate in various ways. Wider transistors increase leakage power, but also enable faster gate transitions by accelerating the discharge of the four intermediate nodes: $\mathrm{N_{HD}}$, $\mathrm{N_{LD}}$, $\mathrm{N_{HI}}$, and $\mathrm{N_{LI}}$. Additionally, the steady-state voltages of these nodes are affected in a non-linear manner by the width of the header and footer transistors, which in turn influences the gate propagation delay, which depends directly on the voltage of intermediate nodes as suggested by (1) and (2).
Given the complexity of these sizing trade-offs, a numerical parametric analysis of the header and footer transistor widths is the best option to explore various sizing configurations.
By extracting leakage power and cell delay across different width combinations, the resulting \ldp can be visualized as a two-dimensional plot relative to header and footer transistor sizes. The optimal size configuration which minimizes the \ldp and ensures the best overall gate performance is then selected from the 2D parameter sweep analysis.


\subsection{Sizing case study}
In the following, we apply the previously defined design and characterization methodology to an \lvt-\flsl inverter and an \lvt-\flsl NAND2 gate. Similar results can be obtained with other combinatorial gate topologies such as AND2, NOR2, and XOR2. Indeed, it can be shown that more complex gates, including series and parallel functional transistors, exhibit an optimal voltage supply ($V_{\mathrm{opt}}$) and must set all of their functional transistors to the minimum allowable width ($W_{\mathrm{min}}$) to minimize \ldp, for the same reasons discussed in \cref{step1_method,step2_method}.

\begin{figure}[h]
    \centering
        \includegraphics[width=0.9\linewidth]{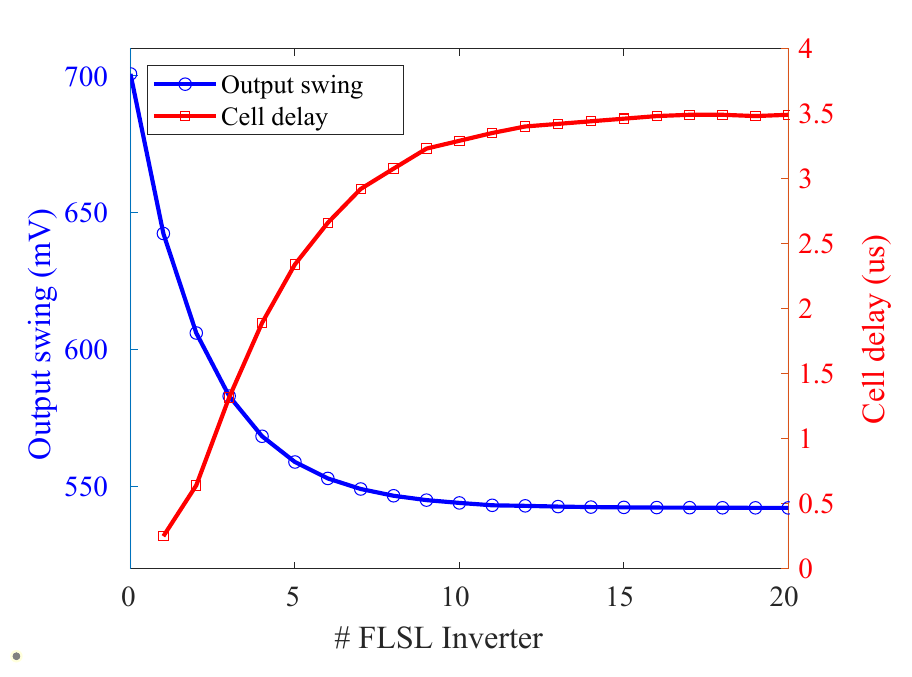}
        \caption{Output swing and cell delay variation on a \SI{20}{}-stage FO4 \flsl inverter chain.}
        \label{fig:swing_degradation}
\end{figure}

All measurements were performed using Cadence Spectre as a spice simulator (version \SI{21.1}{}). To accurately extract the leakage power in the \SI{}{\femto\watt} range, a \SI{50}{}-stage \flsl gate chain is simulated and the total leakage is normalized to obtain the leakage power of a single cell. For delay characterization, the FO4 delay is used as a metric and is extracted following the methodology described in \Cref{delay-characterization}. However, accurate delay estimation in \flsl circuits requires careful isolation from ideal input sources, which can otherwise distort the measured performance. To mitigate this effect, a sufficient number of FO4 \flsl inverters are inserted between the ideal voltage sources and the studied gate. The impact of this isolation is illustrated in \cref{fig:swing_degradation}, which shows the evolution of output swing and propagation delay along a \SI{20}{}-stage FO4 \flsl inverter chain. The propagation delay increases by up to \SI{50}{\times} between the first and \nth{10} inverter stages and only begins to converge after approximately \SI{15}{} stages. Based on this observation, the FO4 delay is extracted from the \nth{20} inverter to ensure sufficient accuracy. Additionally, two extra \flsl inverters are placed after the studied gate to provide a realistic output load.

\begin{figure*}
    \begin{subfigure}{.3\textwidth}
        \centering
        \includegraphics[width=\linewidth]{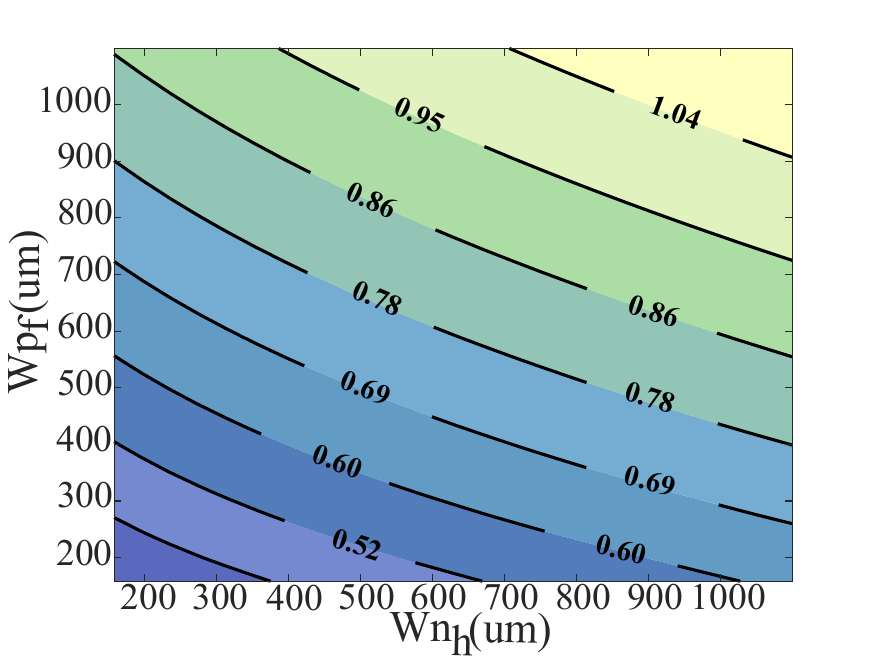}
        \caption{}
        \label{fig:leak_inv_22}
    \end{subfigure}\hfill
    \begin{subfigure}{.3\textwidth}
        \centering
        \includegraphics[width=\linewidth]{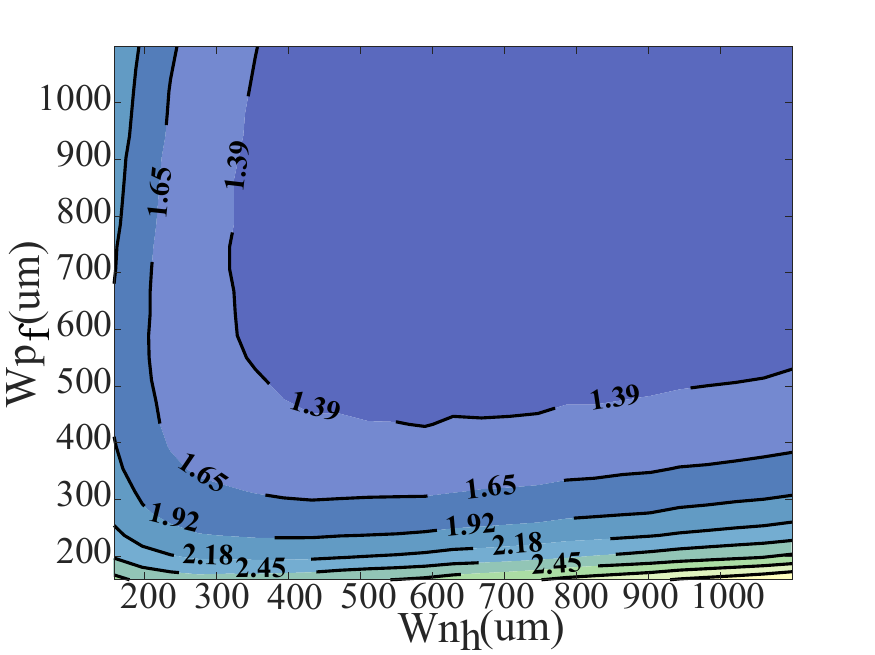}
        \caption{}
        \label{fig:delay_inv_22}
    \end{subfigure}\hfill
    \begin{subfigure}{.3\textwidth}
        \centering
        \includegraphics[width=\linewidth]{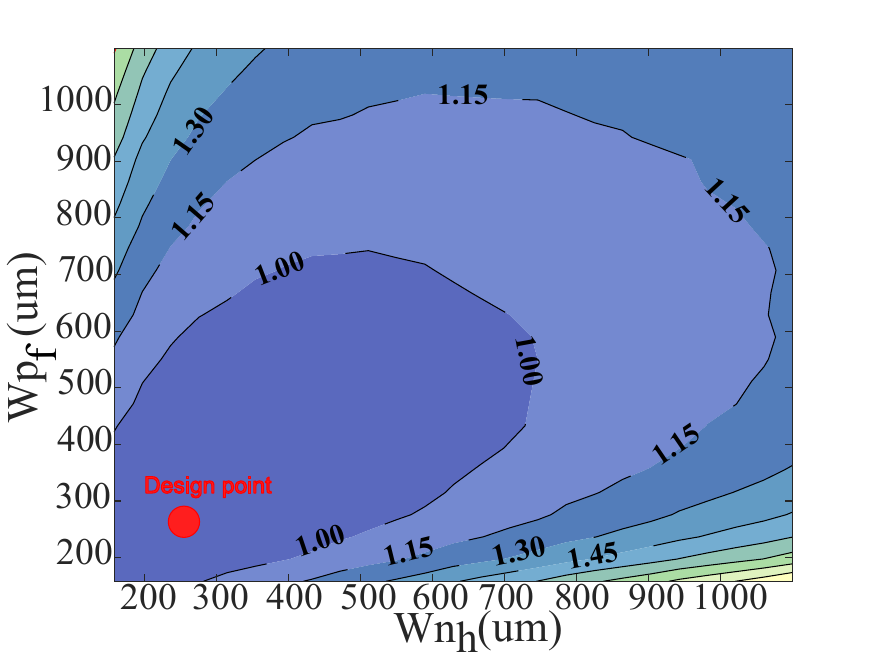}
        \caption{}
        \label{fig:PDP_inv_22}
    \end{subfigure}
    \caption{Leakage power [\si{\pico\watt}] (a), cell delay [\si{\micro\second}] (b), and PDP profile [\si{\atto\joule}] (c) obtained for the \flsl inverter designed with \lvt transistors in a \SI{22}{\nano\meter} \fdsoi process.}
    \label{fig:inv_results}
    \vspace{-2mm}
\end{figure*}

\begin{figure*}
    \begin{subfigure}{.3\textwidth}
        \centering
        \includegraphics[width=\linewidth]{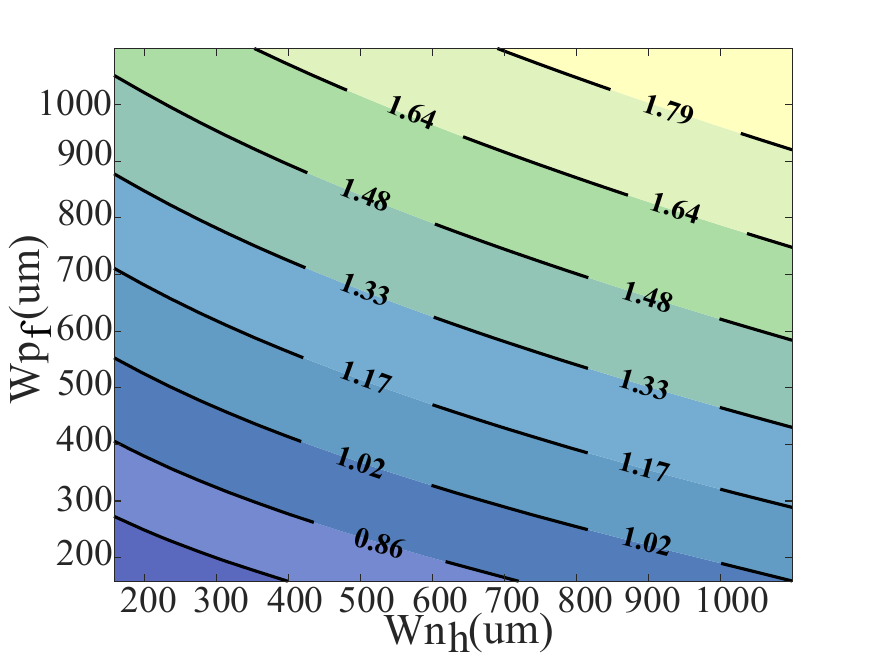}
        \caption{}
        \label{fig:leak_inv_22}
    \end{subfigure}\hfill
    \begin{subfigure}{.3\textwidth}
        \centering
        \includegraphics[width=\linewidth]{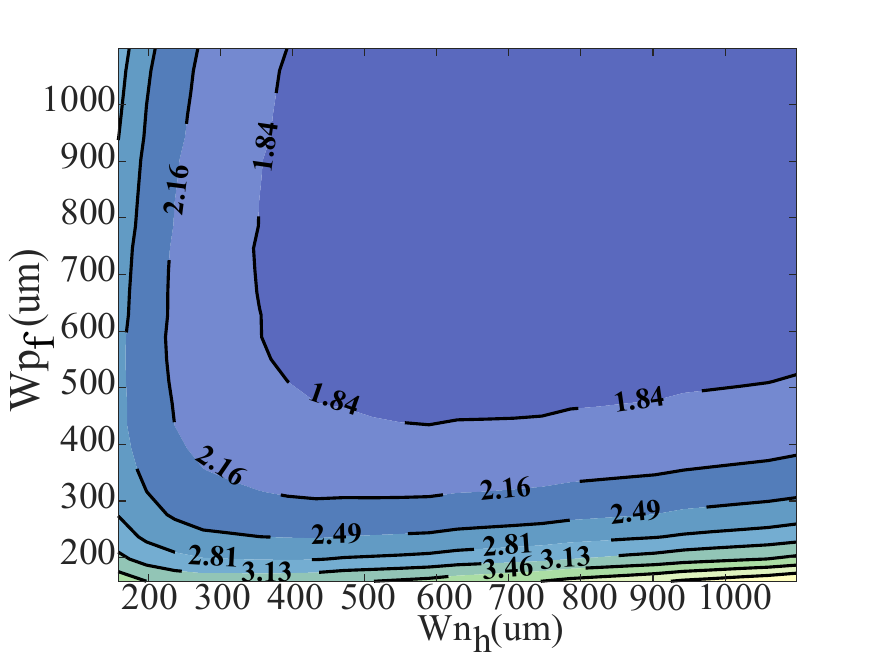}
        \caption{}
        \label{fig:delay_inv_22}
    \end{subfigure}\hfill
    \begin{subfigure}{.3\textwidth}
        \centering
        \includegraphics[width=\linewidth]{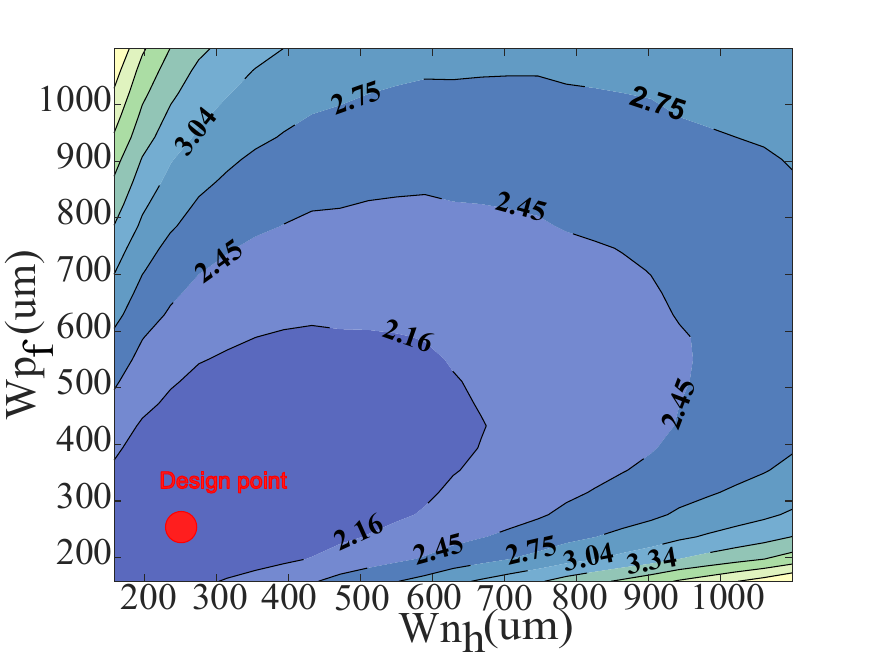}
        \caption{}
        \label{fig:PDP_inv_22}
    \end{subfigure}
    \caption{Leakage power [\si{\pico\watt}] (a), cell delay [\si{\micro\second}] (b), and PDP profile [\si{\atto\joule}] (c) for \flsl NAND2 designed with \lvt transistors in a \SI{22}{\nano\meter} \fdsoi process.}
    \label{fig:nand_results}
    \vspace{-2mm}
\end{figure*}

\cref{fig:inv_results} and \cref{fig:nand_results} report the results extracted from the previous described testbench after applying the final step of the sizing process described in \cref{method} for $V_{\mathrm{opt}}=0.3V$, with the functional transistor widths set to their minimum value. For both inverter and NAND gates, similar trends are observed in leakage power, cell delay, and leakage-power delay product as a function of the header and footer transistor widths. These results indicate that \flsl behaviour and performance are primarily governed by the width ratio of the header and footer transistors rather than by the topology of the pull-up and pull-down networks of the functional devices. The leakage map reflects the relative contributions of the NMOS header in the logic “0” state to the PMOS footer in the logic “1” state, which vary as a function of their respective transistor widths. In the investigated \fdsoi technology, the leakage contributions of the NMOS header and PMOS footer exhibit comparable sensitivity to changes in their width, highlighting a balanced device response.
Similarly, both delay surfaces exhibit a consistent non-linear “V”-shaped dependence on header and footer sizing. This behavior indicates that the optimal delay occurs near the minimum of this curve, corresponding to a configuration in which the PMOS footer width is only slightly larger than that of the NMOS header.
The \ldp trends further confirm this symmetry, as both gates exhibit nearly circular PDP contours with a clearly defined minimum. For the two logic gates under study, the minimum PDP is achieved when both NMOS header and PMOS footer widths are set to \SI{276}{\nano\meter}. Unlike the findings reported by Cerqueira et al. \cite{cerqueira2019femto} for bulk technologies ranging from \SI{180}{\nano\meter} to \SI{28}{\nano\meter}, no significant upsizing of the PMOS footer relative to the NMOS header or specific biasing voltage on the PMOS footer is required to reach optimal performance. This result can be attributed primarily to the properties of the \fdsoi technology, which leads to more balanced and symmetric gate implementations. In particular, the presence of a buried oxide layer and undoped channels reduces body-effect asymmetry, suppresses parasitic PN junction leakage, and limits doping-induced variability, thereby improving device matching compared to bulk technologies.

\subsection{\flsl standard-cell design}

\begin{figure}
    \centering
     \begin{subfigure}{.45\textwidth}
        \centering
        \includegraphics[width=\linewidth]{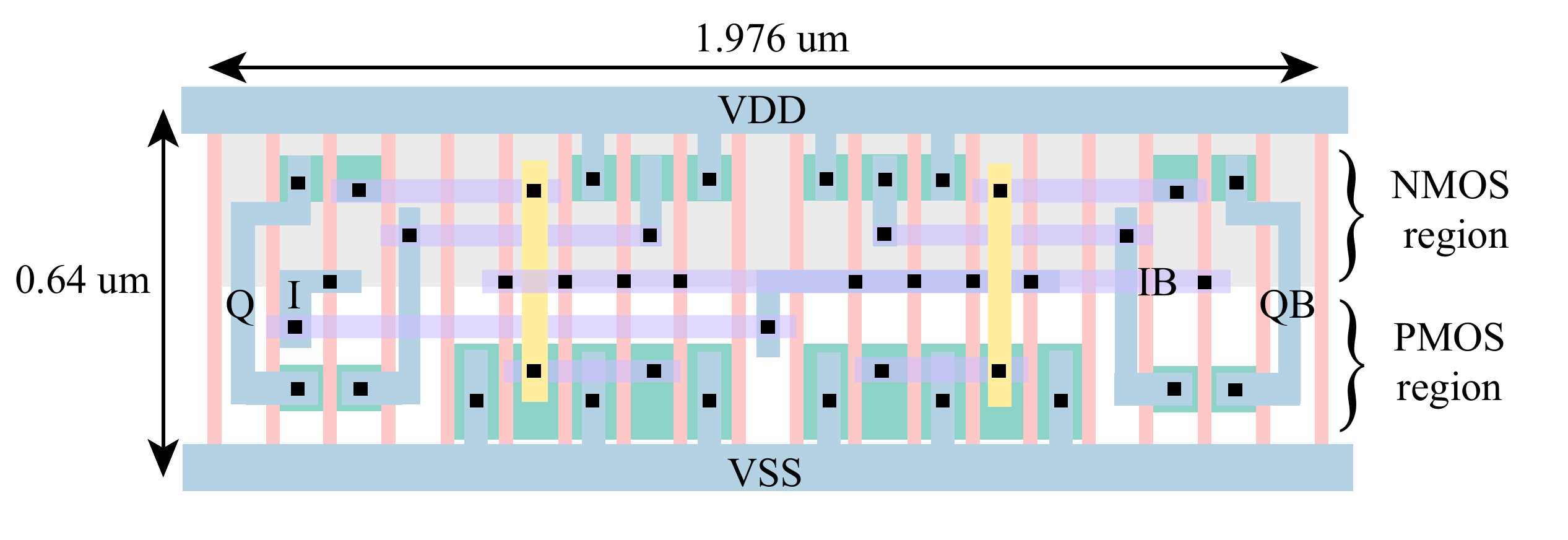}
        \caption{Single-height \flsl-\lvt inverter layout}
        \label{fig:inv_layout}
    \end{subfigure}
    \vspace{0.2cm}  
    \begin{subfigure}{.45\textwidth}
        \centering
        \includegraphics[width=\linewidth]{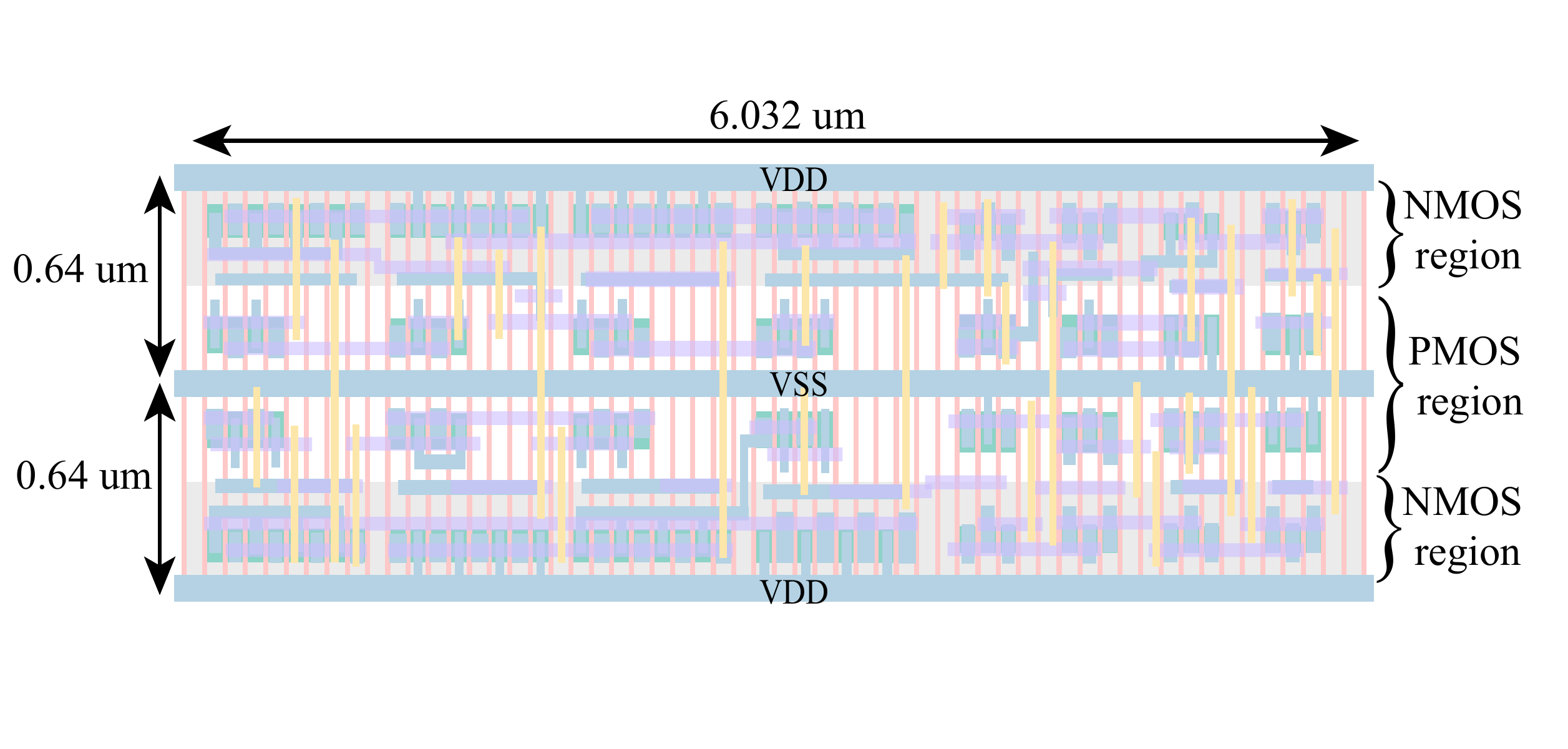}
        \caption{Double-height \flsl-\lvt XOR2 layout}
        \label{fig:xor_layout}
    \end{subfigure}
    \caption{Layout representation of an \flsl-\lvt inverter and an \flsl-\lvt XOR2 gate.}
    \label{fig:flsl-layout}
\end{figure}

To enable the implementation of \flsl circuits within a conventional digital design flow, a dedicated \flsl standard-cell library is required. Accordingly, we developed an \flsl-\lvt library with 14 cells composed of fundamental logic gates, including INV, XOR, MUX, and synchronous-reset D flip-flops. Each cell was sized according to the methodology described in \cref{method} to achieve competitive gate-level performance.
After determining the optimal sizing, all standard cells were laid out with a uniform cell height to ensure compatibility with conventional place-and-route tools. The cell height was selected to match that of a low-power HVT inverter, which will be used later for the clock-tree implementation of the \flsl circuit, as discussed in the next section. The layout of an \flsl-\lvt inverter is shown in \cref{fig:inv_layout}.

Due to the large number of transistors and the differential style of the \flsl logic, layout design is challenging and prone to routing congestion. To mitigate this issue, we adopted a non-conventional device placement strategy, reversing the typical CMOS arrangement by placing NMOS transistors closer to the VDD rail and above the PMOS transistors. This approach significantly reduces routing track length as well as allowing the insertion of \hvt-CMOS inverters and buffers for building the clock-tree as explained later in \cref{sec:impl}. For complex cells such as XOR2 and XNOR2 gates, which remain congested even after this optimization, a double-height layout (i.e., twice the reference cell height) is employed to ensure proper placement and routing as shown in \cref{fig:xor_layout}.

\subsection{Circuit implementation}
\label{sec:impl}

In this section, we present the proposed design flow for circuit implementation in \flsl logic, summarized in \cref{fig:flsl_flow}. To enable fast and efficient \flsl circuit design, the flow takes advantage of conventional synthesis and place-and-route tools. The circuit is first described using a standard hardware description language (HDL). This description remains fully agnostic to the final \flsl implementation and requires no explicit modeling of differential inputs or outputs.

Next, conventional synthesis is performed using Synopsys Design Compiler® 2020.09 under the constraint of using only the restricted subset of the designed \flsl standard cells. Also, to minimize leakage and gate count, relaxed timing constraints are chosen. Technology mapping during synthesis is done against the standard CMOS library. The resulting post-synthesis netlist is then translated into an \flsl netlist using a Python script, which replaces CMOS cells with their \flsl counterparts and automatically introduces the inverted signal path that is required for differential operation. Additionally, a set of graph-based optimizations specific to the \flsl differential style is applied, such as selective removal of redundant inverters while preserving essential buffering structures.

Physical implementation is carried out using Cadence Innovus® 20.1. Due to the low operating frequency of \flsl logic cells, timing analysis is omitted, and the tool is primarily used for placement. To ensure the correct operation of sequential elements, a clock tree based on regular \hvt-CMOS inverters is implemented. This choice is motivated by both functional and physical considerations. First, the three-orders-of-magnitude speed difference between CMOS and \flsl prevents hold violations, ensuring correct operation of low-frequency circuits. Second, to minimize the leakage overhead introduced by the clock tree, \hvt or \uhvt devices are required. Although \uhvt transistors offer lower static leakage, their site width is incompatible with the \lvt devices used in \flsl standard cells. In contrast, \hvt transistors provide a compatible layout, enabling area-efficient integration. Although \hvt and \lvt transistors rely on different well structures (regular versus flipped wells), the reversed layout of \flsl cells enables direct connection of N-well and P-well regions between CMOS and \flsl gates, allowing seamless integration of the \hvt-CMOS inverters during clock-tree insertion.

Finally, the post-layout Verilog netlist generated by Innovus is converted into a SPICE netlist using Calibre v2lvs from Siemens. The resulting netlist is then simulated with the fast SPICE simulator CustomSim 2021.09 from Synopsys to perform final functional, dynamic timing and power analysis.

\section{Silicon Validation}
\label{sec:silicon}

Using the previously described \flsl standard-cell library and adapted digital design flow, two representative \flsl circuits for \ao processing, an 8-bit, 8-taps FIR filter and \aes cryptographic core \cite{nationalinstituteofstandardsandtechnologyusAdvancedEncryptionStandard2001}, are implemented in \SI{22}{\nano\meter} \fdsoi technology on an experimental chip called HEEPnosis. For future comparisons, equivalent ultra-low-power CMOS implementations are also implemented using ultra low-power \hvt and \uhvt standard-cell libraries.

\subsection{Chip prototype: HEEPnosis}

\begin{figure}
    \centering
     \begin{subfigure}{.45\textwidth}
        \centering
        \includegraphics[width=\linewidth]{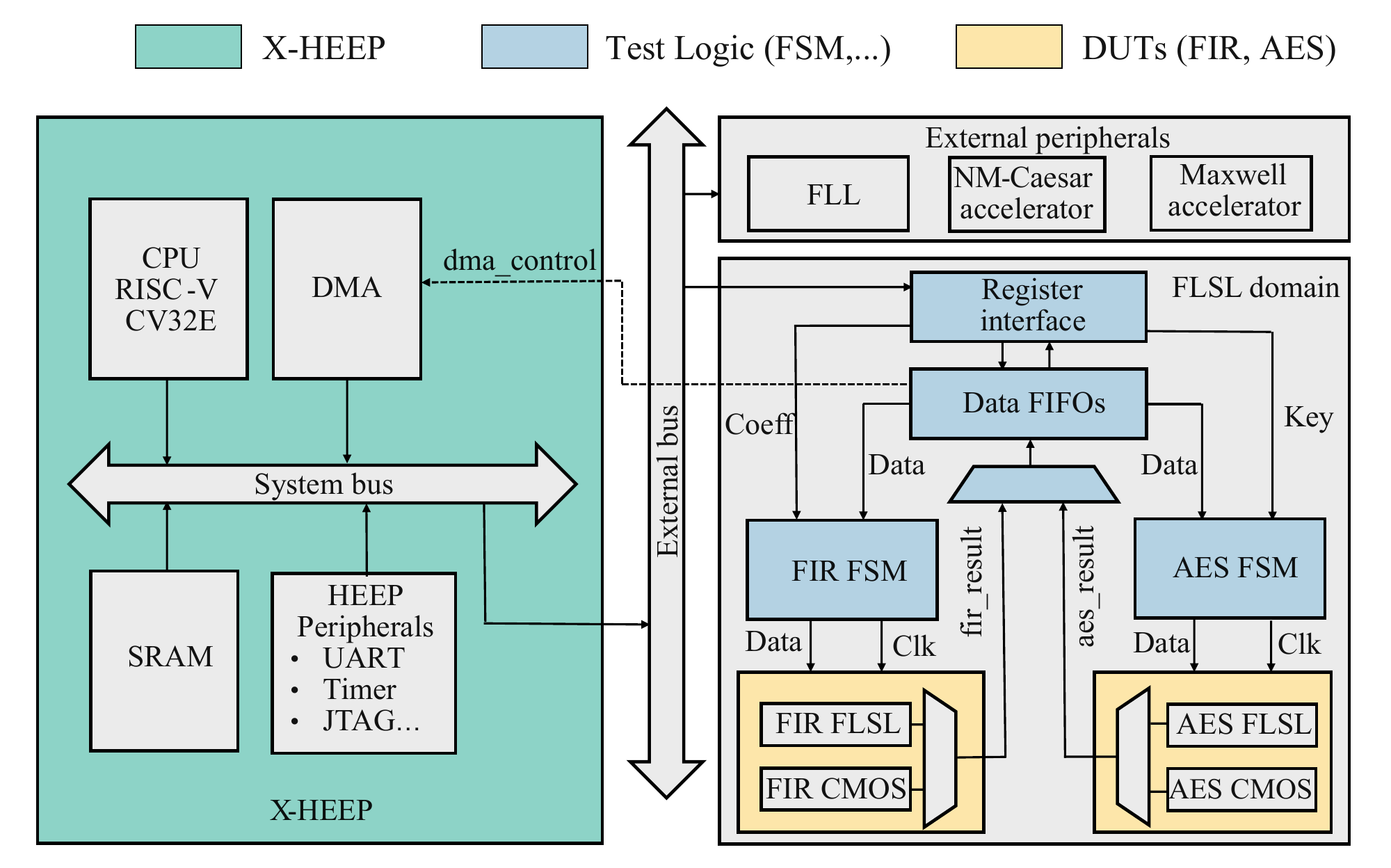}
        \caption{}
        \label{fig:heepnosis_bd}
    \end{subfigure}
    \vspace{0.2cm}  
    \begin{subfigure}{.45\textwidth}
        \centering
        \includegraphics[width=\linewidth]{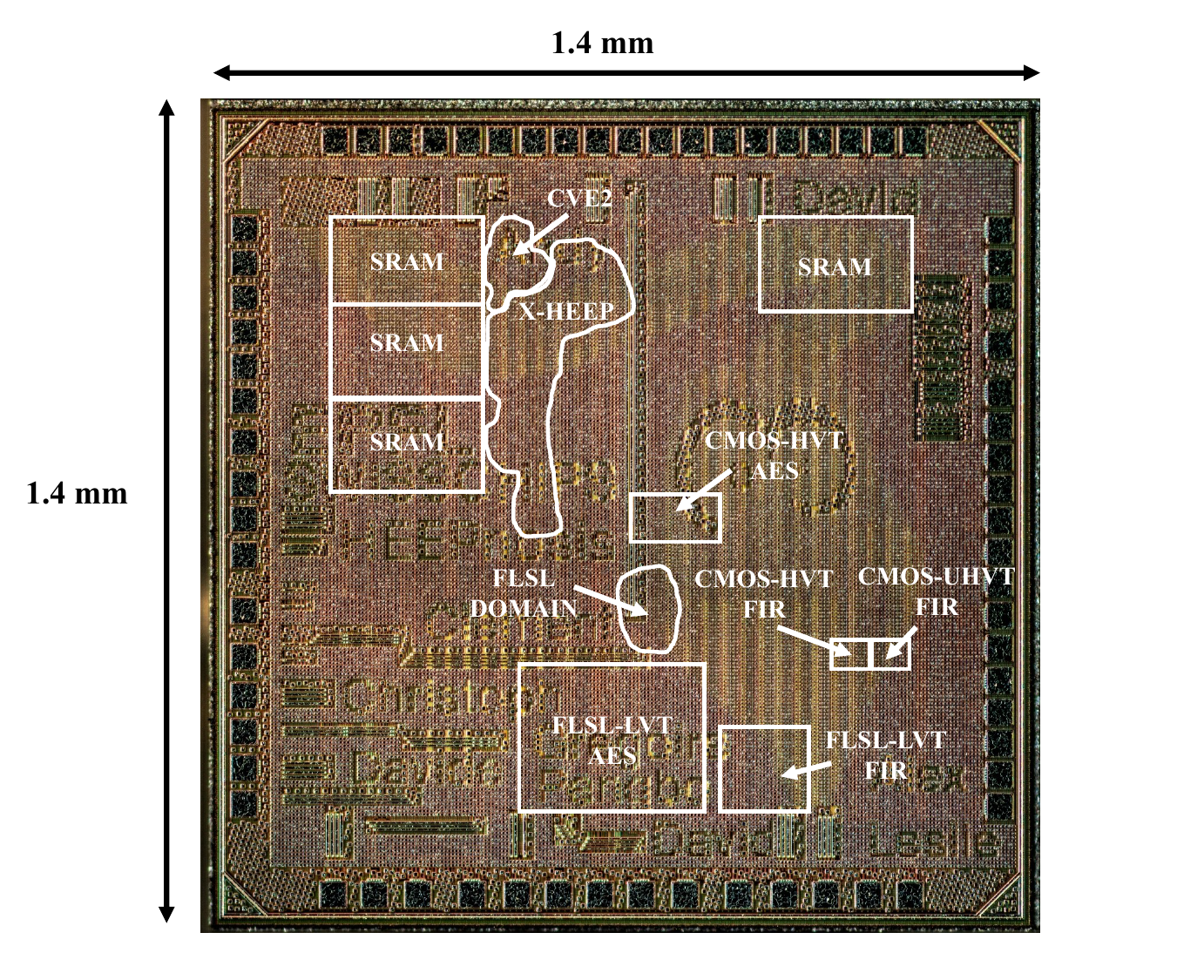}
        \caption{}
        \label{fig:heepnosis_fp}
    \end{subfigure}
    \caption{Block diagram (a) and microphotograph (b) of the testing chip HEEPnosis.}
    \label{fig:HEEPnosis}
\end{figure}

To validate and characterize the \flsl logic blocks, the \flsl circuits are integrated into HEEPnosis, a chip build with \hvt standard CMOS logic and based on X-HEEP \cite{machetti2024xheep}, an open-source, configurable low-power RISC-V microcontroller. A block diagram and the chip layout are shown in \cref{fig:HEEPnosis}. The microcontroller manages data transfers between the \flsl domain and SRAM banks via an integrated DMA peripheral. Programmable finite-state machines (FSMs) are integrated into the peripheral subsystem of X-HEEP to drive and stimulate the FIR filters and \aes blocks, i.e., the devices under test (DUTs). The FSMs are used to generate independent and programmable clock signals for \flsl logic, introducing controlled delays in the input signals with respect to the generated clock to mitigate potential timing violations at the interface, and introducing controlled delays in the DUT.output sampling.
The X-HEEP microcontroller, together with its memories, peripherals, and FSMs to drive the DUTs, is synthesized, placed, and routed following the standard flow and implemented using the CMOS HVT library. Each DUT has its own voltage dedicated pins to isolate its power consumption from the rest of the chip for accurate measurements. In addition, level-shifters are added at the boundaries of the DUTs to characterize them for different voltages independently of the supply voltage of X-HEEP.
After each DUT was implemented following the methodology described in \cref{sec:impl}, their liberty timing file (lib) and the library exchange format (lef) are used for integrating them into X-HEEP. The lib files of \flsl DUTs do not contain meaningful timing information, thus timing is validated using analog-mixed-signal (AMS) verification, co-simulating the back-annotated post-layout Verilog netlist of X-HEEP, which relies on the PDK models of memories and standard cells, with the spice netlist of the DUTs using VCS 2022 and CustomSim 2021.09 from Synopsys.

\begin{figure}[htb]
    \centering
    \includegraphics[width=\linewidth]{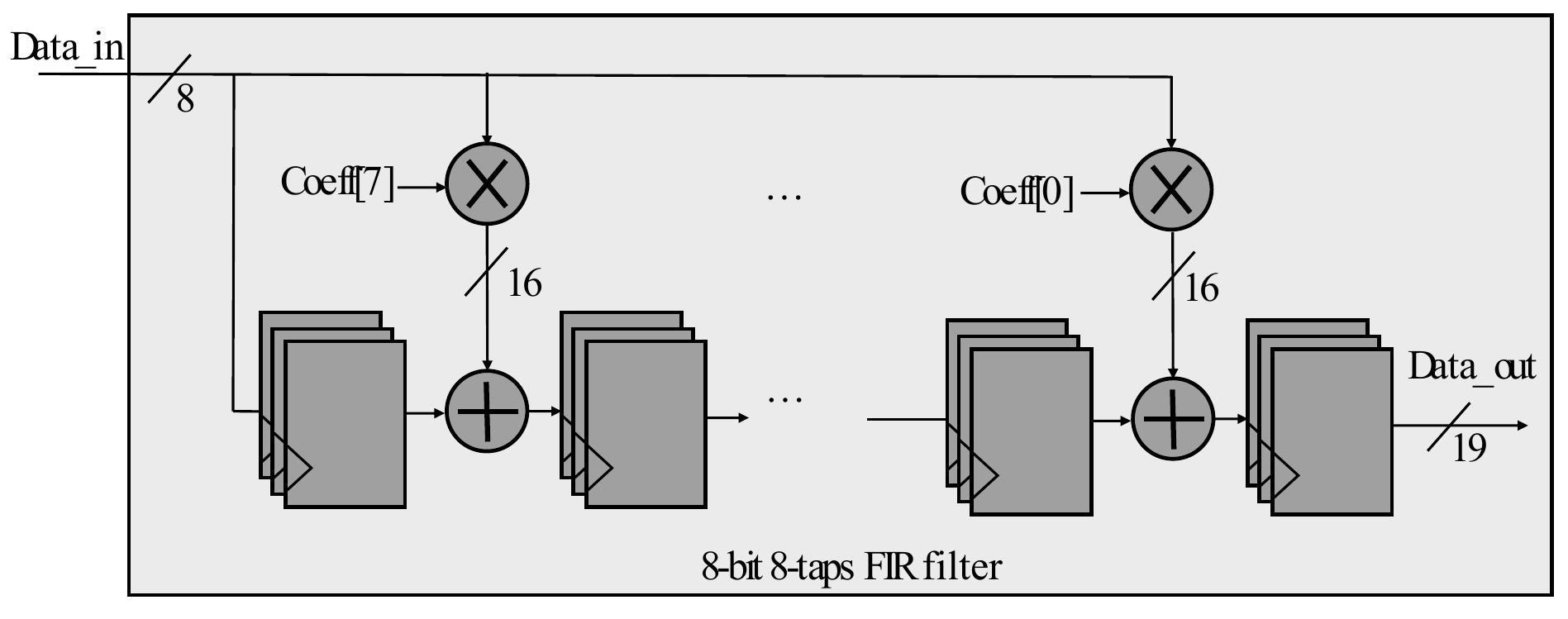}
    \caption{Block diagram of implemented FIR filter.}
    \label{fig:block-diagram-FIR}
\end{figure}

\begin{figure}
    \centering
    \includegraphics[width=\linewidth]{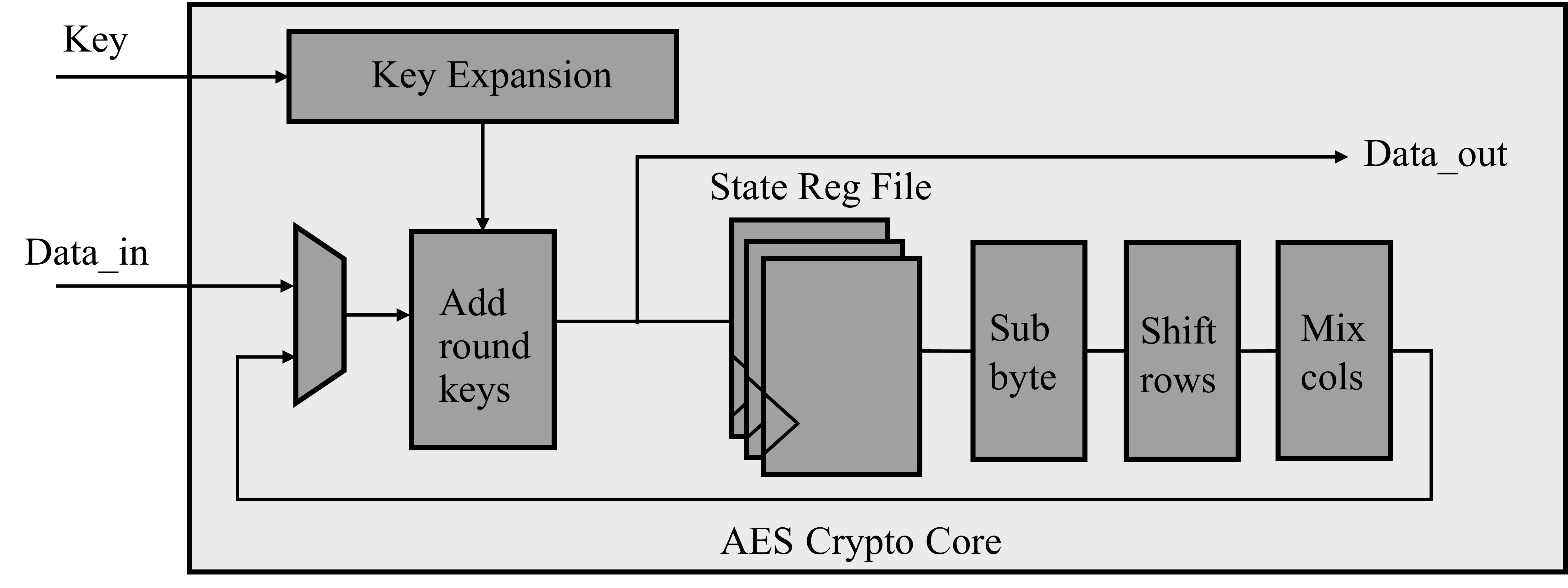}
    \caption{Block diagram of implemented \aes block.}
    \label{fig:aes}
\end{figure}

The implemented FIR filter follows a transposed architecture, as shown in \cref{fig:block-diagram-FIR}, and is composed of 3,400 logic gates. The \flsl version, implemented using \lvt transistors, occupies a silicon area of 21,291\,$\mu m^2$, corresponding to a silicon area overhead of \SI{6.3}{\times} compared to equivalent \uhvt and \hvt CMOS implementations. \\
As illustrated in Figure~\ref{fig:aes}, the \aes hardware architecture employs a standard iterative structure. The design accepts a plaintext input alongside an initial key. A key expansion module processes this initial key to generate the required round keys for each encryption iteration. Composed of 10,650 logic gates, the \flsl version, implemented as well with \lvt transistor occupies a silicon area of 
81,826\,$\mu m^2$, which corresponds to a silicon area overhead of \SI{6.7}{\times} compared to an equivalent \hvt CMOS implementation.

\subsection{\flsl power consumption}

During functional and power measurement tests, the temperature of the chip was controlled using a temperature controller, and a pico-precision probe was used to perform accurate leakage measurements. For both FIR and \aes architectures, equivalent \hvt and \uhvt CMOS implementations with high \rbb are used as baseline, while the \flsl variant is evaluated under three body-biasing conditions: zero body bias (0V, 0V), medium \fbb  (-1V, 1V), and high \fbb (-1.6V, 1.6V).

\subsubsection{8-bit 8-TAP FIR filters} 
\begin{figure}
    \centering
     \begin{subfigure}{.4\textwidth}
        \centering
        \includegraphics[width=\linewidth]{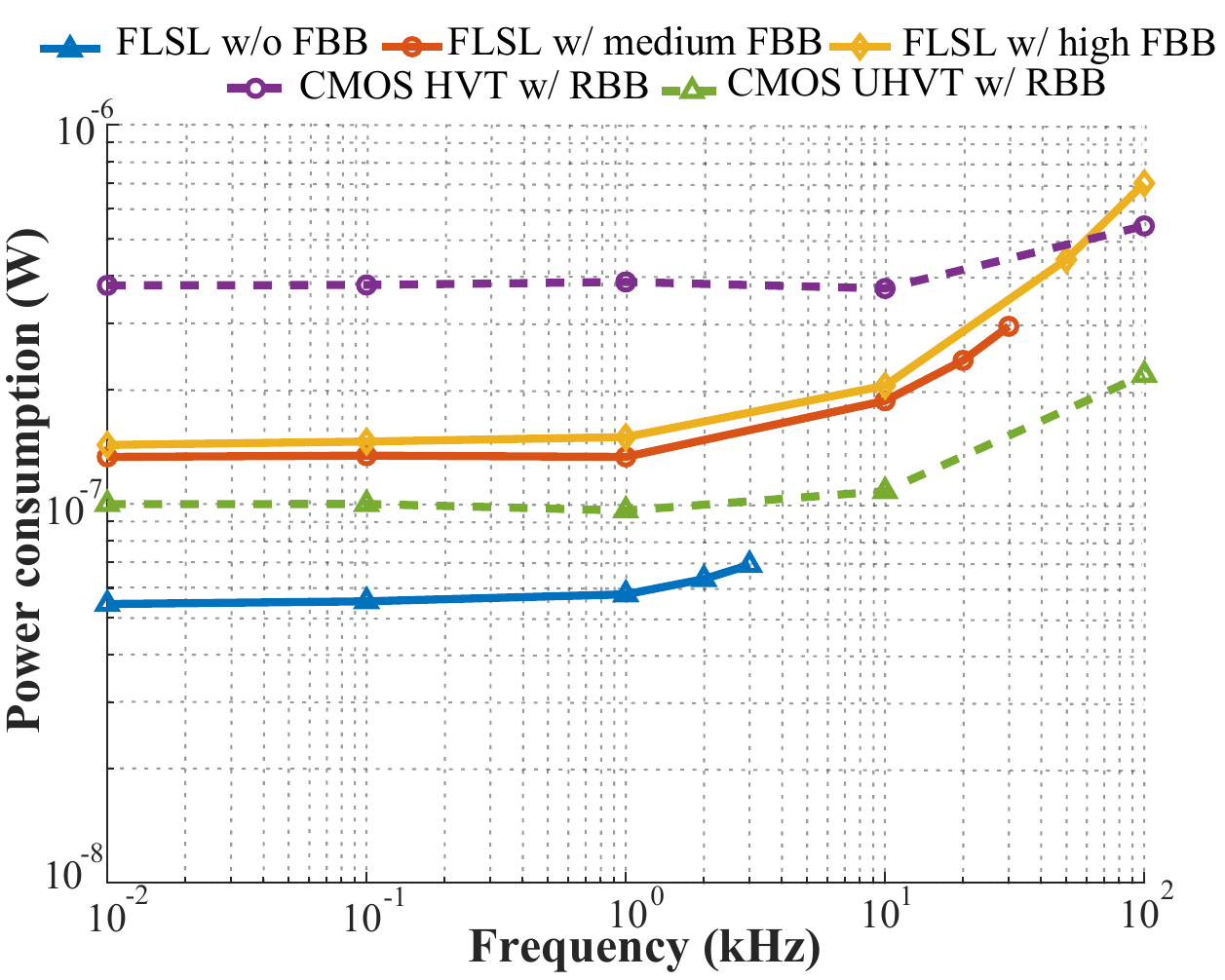}
        \caption{Nominal supply voltage (0.8V)}
        \label{fig:0p8}
    \end{subfigure}
    \vspace{0.2cm}  
    \begin{subfigure}{.4\textwidth}
        \centering
        \includegraphics[width=\linewidth]{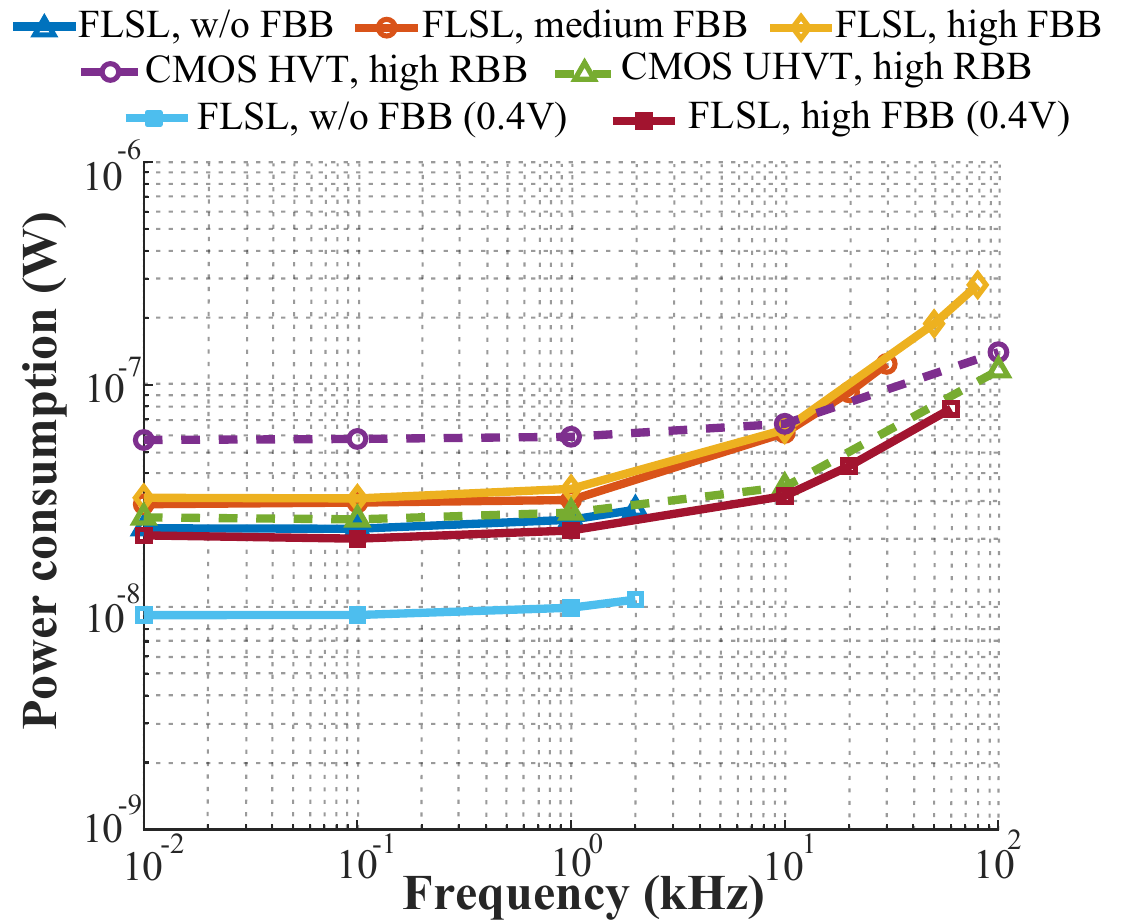}
        \caption{Reduced supply voltage (0.6V or 0.4V)}
        \label{fig:0p6}
    \end{subfigure}
    \caption{ \flsl-\lvt FIR filter power consumption vs frequency at \SI{25}{\degreeCelsius}.}
    \label{fig:flsl_power}
\end{figure}

\cref{fig:flsl_power} presents the static leakage and total power consumption of the \flsl-\lvt FIR filter at nominal supply voltage (i.e., $V_{DD}=0.8 V$) as well as under reduced-voltage operation for three body biasing conditions. At nominal voltage, the \flsl filter achieves a leakage power as low as \SI{54.5}{\nano\watt}, corresponding to a \SI{6.9}{\times} and \SI{1.83}{\times} reduction compared to equivalent FIR implementations based on \hvt and \uhvt CMOS standard-cell libraries, respectively, both operating under \rbb.

Thus, without \fbb, \flsl achieves the lowest energy consumption among all FIR implementations, but only at relatively low operating frequencies (up to \SI{3}{\kilo\hertz}). Applying \fbb extends the operating range to several tens of kilohertz, at the expense of increased static leakage and dynamic power. Although the highly forward biased configuration enables energy savings up to \SI{60}{\kilo\hertz} compared to an equivalent \hvt-CMOS implementation, the implementation with \uhvt-CMOS logic remains more competitive over this frequency range at nominal voltage. Consequently, \flsl-based always-on circuits are more energy-efficient for applications operating up to \SI{3}{\kilo\hertz}, whereas \uhvt-CMOS implementations provide better energy efficiency at higher frequencies.

\cref{fig:0p6} presents the results obtained for the FIR implementation at reduced supply voltage for different body biasing configurations. At \SI{0.6}{\volt}, similar results compared to nominal operation are obtained. Although the \flsl filter without body biasing achieves the lowest leakage (i.e. \SI{22.6}{\nano\watt}), the relative leakage advantage of \flsl over conventional CMOS logic diminishes as the supply voltage decreases. This behaviour can be explained by the operating principle of \flsl, whose leakage reduction relies on the strength of the super-cutoff effect induced by the additional header NMOS and footer PMOS transistors. The effectiveness of this mechanism is directly related to the magnitude of the negative gate bias applied to these devices, which scales with the supply voltage. Consequently, lowering the supply voltage reduces the achievable gate overdrive that enforces the super-cutoff condition, thereby weakening the leakage suppression capability of \flsl and narrowing its benefit relative to conventional CMOS circuits.\\

\subsubsection{\aes Accelerator}

\begin{figure}
    \centering
     \begin{subfigure}{.4\textwidth}
        \centering
        \includegraphics[width=\linewidth]{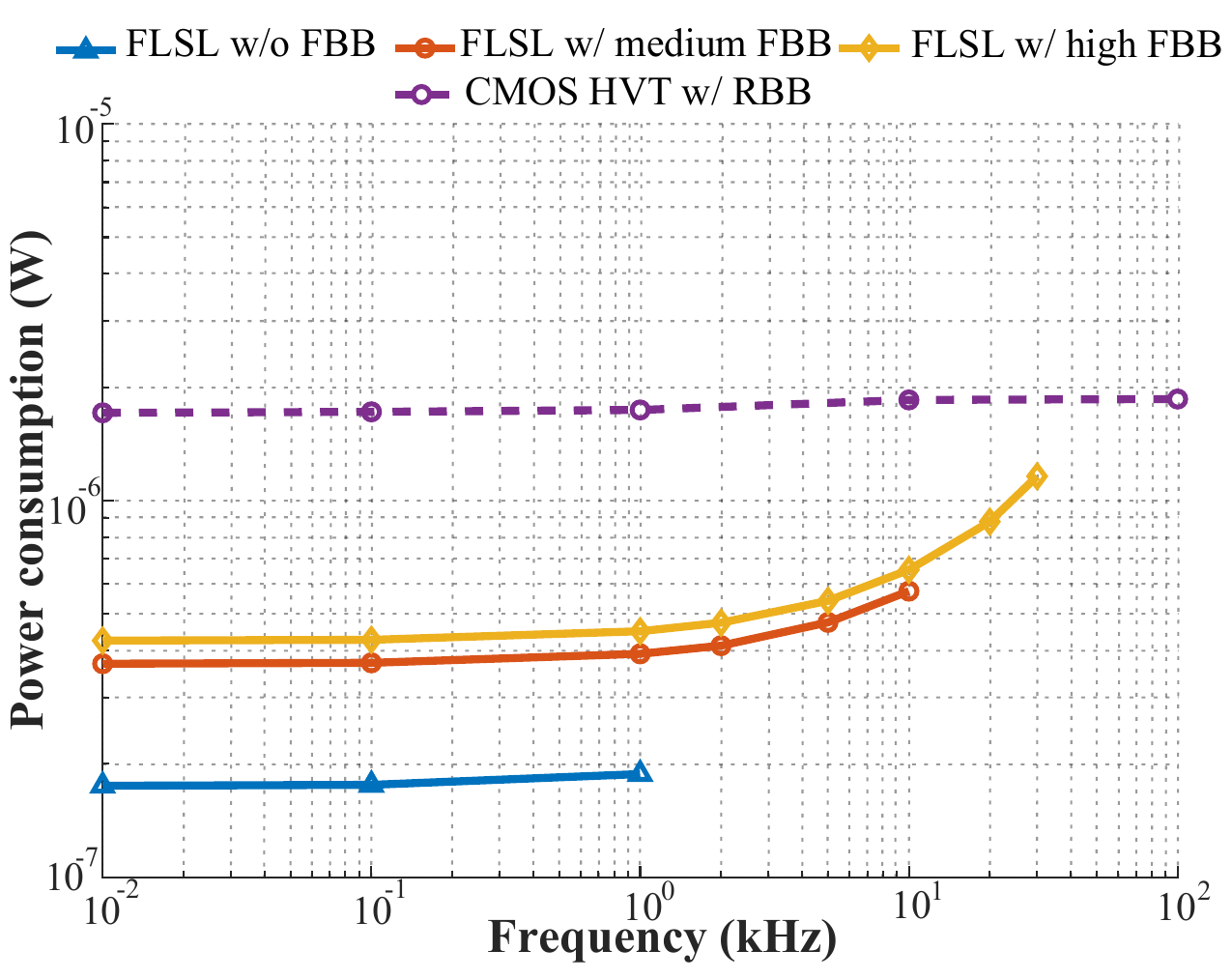}
        \caption{Nominal supply voltage (0.8V)}
        \label{fig:aes0p8}
    \end{subfigure}
    \vspace{0.2cm}  
    \begin{subfigure}{.4\textwidth}
        \centering
        \includegraphics[width=\linewidth]{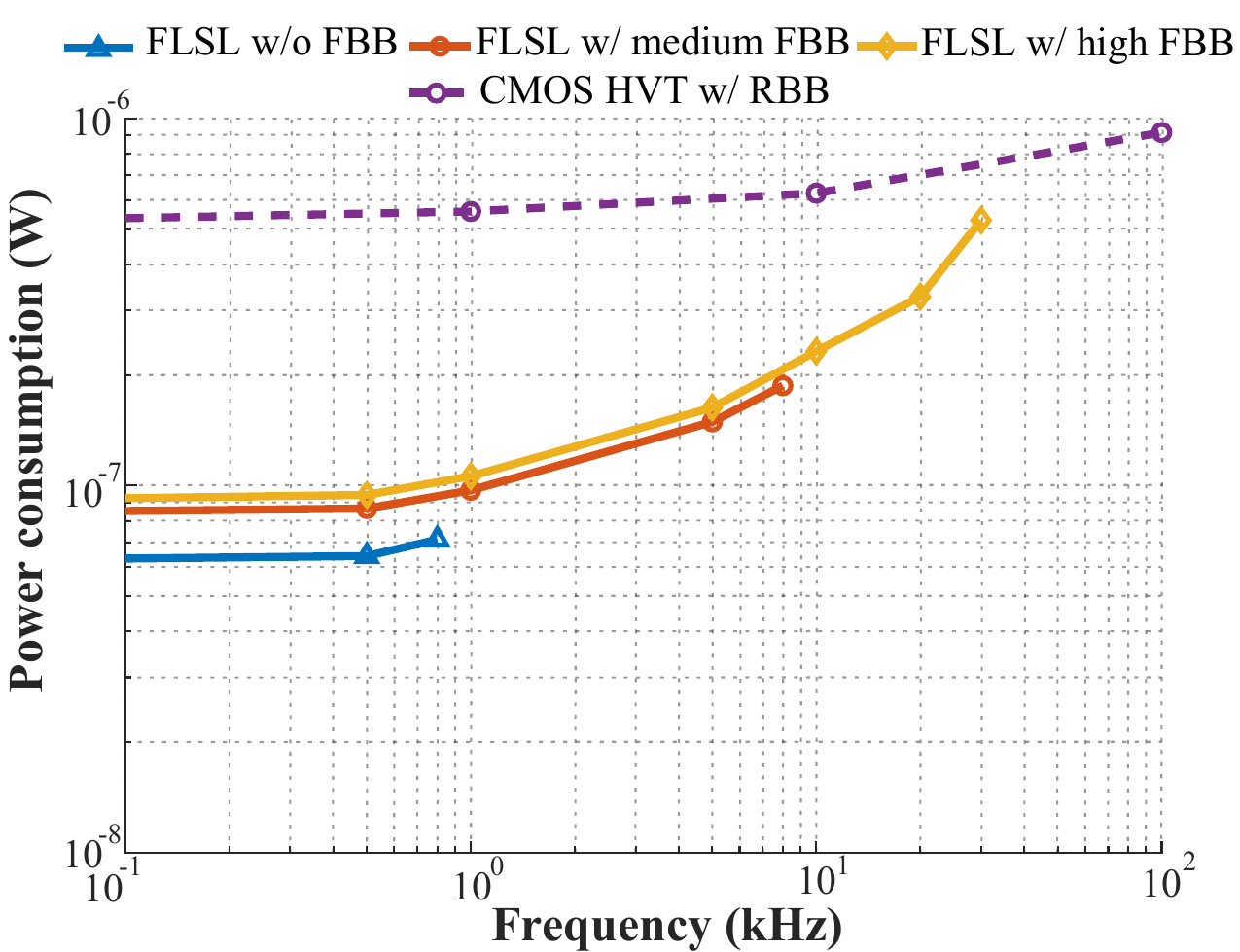}
        \caption{Reduced Supply voltage (0.6V)}
        \label{fig:aes0p6}
    \end{subfigure}
    \caption{ \flsl-\lvt \aes power consumption vs frequency at \SI{25}{\degreeCelsius}.}
    \label{fig:flsl_aes_power}
\end{figure}
\cref{fig:flsl_aes_power} reports silicon measurement results of static leakage and power consumption at $V_{DD} = \SI{0.8}{\volt}$ and \SI{0.6}{\volt} for the \aes core. Scaling the supply voltage from \SI{0.8}{\volt} down to \SI{0.6}{\volt} results in a minor decrease in the maximum operating frequency ($f_{max}$), shifting from \SI{33}{\kilo \hertz} to \SI{30}{\kilo \hertz}. Within this operational frequency range, the \flsl implementation consumes overall less power compared to the standard \hvt-CMOS baseline, achieving up to \SI{9.8}{\times} and \SI{8.5}{\times} leakage reduction at \SI{0.8}{\volt} and \SI{0.6}{\volt} respectively. Similarly to the \flsl FIR filters, it can be notice that \flsl reduction mechanism diminishes with voltage scaling and \flsl logic exhibits a steeper increase in dynamic power consumption compared to its \hvt-CMOS variant over frequency due to the higher transistor count and associated higher parasitic capacitance per gate. Ultimately, \flsl architectures successfully reduces total power consumption for low-frequency operations while conventional CMOS architectures achieve higher energy efficiency at higher frequencies.

\subsection{\flsl robustness}

\begin{figure}
    \centering
        \includegraphics[width=0.9\linewidth]{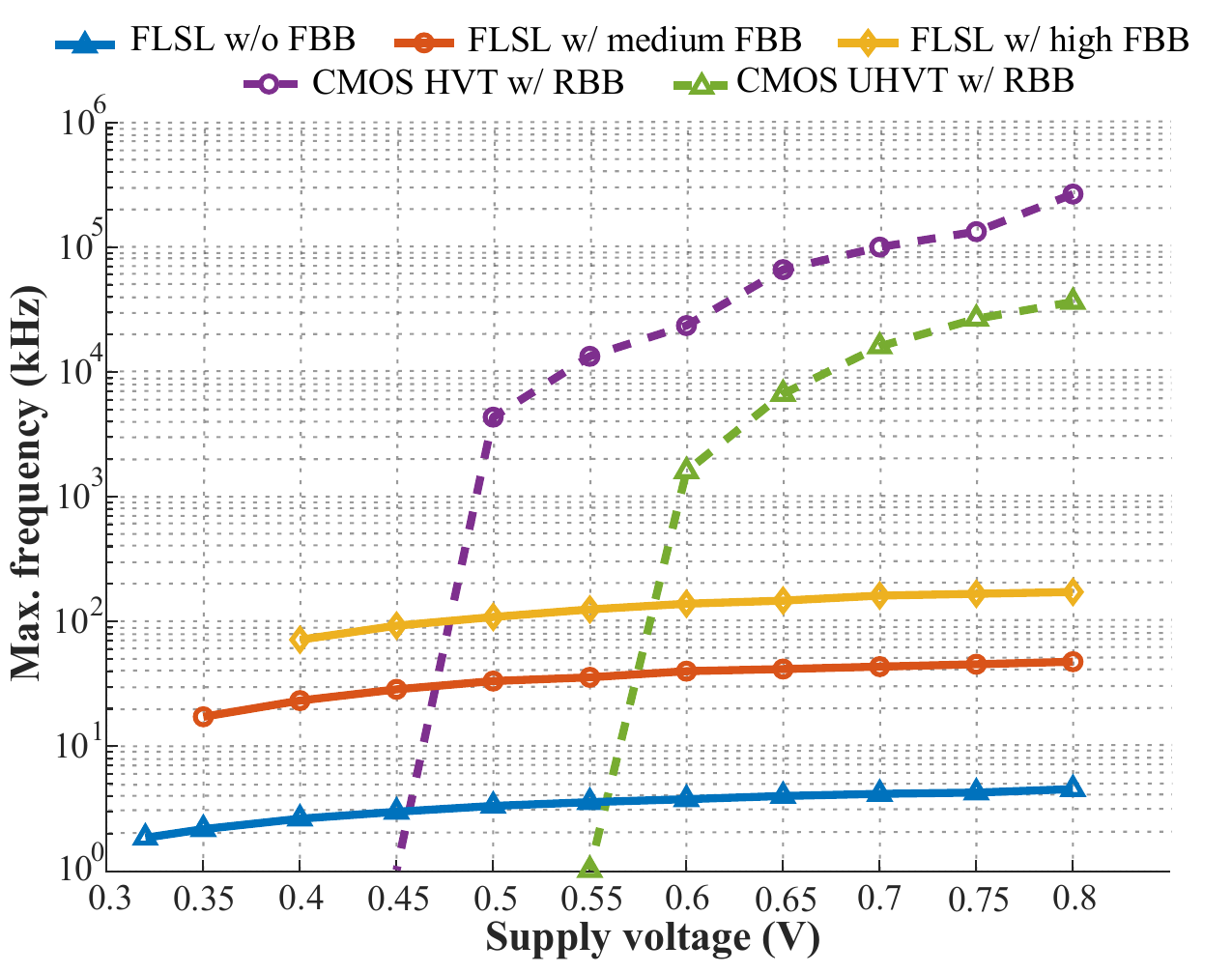}
        \caption{Maximum operating frequency versus power supply measured on silicon for an 8-bit 8-TAP FIR filter.}
        \label{fig:LSL_freq}
\end{figure}

After assessing \flsl performance, its robustness to \pvt variations should also be evaluated. \cref{fig:LSL_freq} presents the measured maximum operating frequency of the 8-bit, 8-tap FIR filter fabricated in \SI{22}{\nano\meter} \fdsoi technology using both \flsl and ultra-low-power CMOS logic styles. As the supply voltage is reduced from \SI{0.8}{\volt} to \SI{0.6}{\volt}, the maximum operating frequency of the two CMOS implementations decreases by more than one order of magnitude, whereas the \flsl-based design exhibits only a modest degradation. When the supply voltage is further lowered, the performance of the CMOS filters drops sharply, preventing correct operation below \SI{0.5}{\volt}. In contrast, the \flsl implementation maintains its lower but significantly more stable performance degradation trend and remains fully functional down to \SI{0.4}{\volt}, where it can still process data above \SI{100}{\kilo\hertz} when high \fbb is applied. At this supply voltage and up to \SI{2}{\kilo\hertz}, the unbiased \flsl filter consumes twice less energy than the \uhvt-CMOS variant operating at \SI{0.6}{\volt}. When \fbb is applied, the \flsl filter consumes lower or similar energy as the \uhvt-CMOS variant operating at \SI{0.6}{\volt}, providing therefore an alternative to \uhvt devices which are not fully physically compatible with \lvt and \hvt devices.

This improved voltage scalability can be attributed to the intrinsic switching mechanisms of \flsl based on leakage currents, which preserve effective switching conditions even at reduced supply voltages. As a result, \flsl logic demonstrates superior robustness in low-voltage operations compared to conventional CMOS implementations. \\

\begin{table*}[th]
\centering
\begin{tabular}{lcccccc}
\toprule
 & \multicolumn{2}{c}{Leakage power (0.8V)} 
 & \multicolumn{2}{c}{Leakage power (0.6V)} 
 & \multicolumn{2}{c}{Temperature coefficient (85--25)} \\
\cmidrule(lr){2-3} \cmidrule(lr){4-5} \cmidrule(lr){6-7}
 & Average (nW) & Spread (\%) & Average  (nW) & Spread  (\%) & VDD = 0.8V & VDD = 0.6V \\
\midrule
\flsl (no biasing) & \textbf{64.4} & $\pm$14.3 & \textbf{26.7} & $\pm$13.7 & 6.7 & 6.0 \\
\flsl (High FBB)   & 152.8 & $\pm$6.8  & 32.8 & \textbf{$\pm$10.3} & \textbf{3.3} & \textbf{5.6} \\
CMOS \hvt (RBB)    & 383.8 & \textbf{$\pm$2.4} & 67.4 & $\pm$17.3 & 4.5  & 7.4 \\
CMOS \uhvt (RBB)   & 101.4 & $\pm$22.0 & 32.2 & $\pm$29.8 & 6.8  & 9.1 \\
\bottomrule
\end{tabular}
\caption{Leakage variation under different voltage and temperature conditions measured on 5 chips.}
\label{tab:leakage}
\end{table*}

To further evaluate the robustness of the \flsl approach, the variability of the leakage was also characterized according to process and temperature variations. Process variability was assessed using measurements from multiple fabricated chips, while temperature dependence was evaluated using a controlled thermal setup. This analysis allows assessing the sensitivity of \flsl to manufacturing dispersion and environmental variations, providing information on its reliability and stability under realistic operating conditions.
Leakage measurements of both \flsl and CMOS FIR filter implementations were performed on five chips at \SI{25}{\degreeCelsius} and \SI{85}{\degreeCelsius}, and the results are reported in \cref{tab:leakage}. In \SI{25}{\degreeCelsius}, the \flsl design without forward body biasing exhibits lower average leakage than any CMOS variant that employs \rbb, at both nominal and reduced supply voltages. In particular, the \flsl implementation achieves \SI{64.4}{\nano\watt} at nominal voltage, corresponding to a reduction of \SI{6}{\times} and \SI{1.6}{\times} compared to the \hvt and \uhvt CMOS version. When \fbb is applied, leakage is reduced by \SI{2.5}{\times} relatively to the \hvt CMOS design under \rbb, although it remains \SI{1.2}{\times} higher than the \uhvt CMOS counterpart.
To assess robustness at elevated temperature, a temperature coefficient was defined as the ratio of leakage measured at \SI{85}{\degreeCelsius} to that at \SI{25}{\degreeCelsius}; lower values therefore indicate greater thermal stability. Under this metric, \flsl demonstrates comparable or smaller temperature coefficients than CMOS implementations, indicating improved resilience to temperature variations.

Overall, these results show that \flsl exhibits comparable or lower sensitivity to process, supply voltage, and temperature variations than CMOS, making it a strong candidate for applications requiring stable operation across varying environmental conditions.

\section{Conclusion}
\label{sec:conclusion}

\noindent Towards the development of the next generation of ultra-low-power always-on circuits, this article has presented the design, characterization, and implementation of \flsl circuits in a \SI{22}{\nano\meter} \fdsoi technology. We have demonstrated that \fdsoi enables significant leakage reduction and improved performance tuning via body biasing compared to low-power CMOS logic, thereby overcoming the limitations of \flsl in advanced technology nodes. A comprehensive methodology for \flsl design was introduced, encompassing device-level characterization, systematic transistor sizing, and circuit-level implementation, facilitating efficient design-space exploration and practical deployment. Silicon measurements on complete \flsl circuits validate the proposed approach, showing up to \SI{9.8}{\times} and \SI{1.83}{\times} leakage reduction compared to state-of-the-art ultra-low-power \hvt and \uhvt CMOS design kits, respectively, in addition to exhibit comparable or lower sensitivity to \pvt variations. These results confirm that the combination of \fdsoi technology and \flsl logic provides a compelling path toward energy-efficient \ao systems, paving the way for future edge AI and IoT applications where standby leakage power is a critical constraint.

\section{Acknowledgements}
This work was supported in part by the Swiss State Secretariat for Education, Research, and Innovation (SERI) through the SwissChips research project.



\begin{acronym}
    \acro{ai}[AI]{Artificial Intelligence}
    \acro{alu}[ALU]{Arithmetic Logic Unit}
    \acro{ann}[ANN]{Artificial Neural Network}
    \acro{api}[API]{Application Programming Interface}
    \acro{asic}[ASIC]{Application-Specific Integrated Circuit}
    \acro{cgra}[CGRA]{Coarse-Grained Reconfigurable Architecture}
    \acro{cim}[CIM]{Compute-In-Memory}
    \acro{cnn}[CNN]{Convolutional Neural Network}
    \acro{cpu}[CPU]{Central Processing Unit}
    \acro{csr}[CSR]{Control and Status Register}
    \acro{dma}[DMA]{Direct Memory Access}
    \acro{dnn}[DNN]{Deep Neural Network}
    \acro{dram}[DRAM]{Dynamic Random Access Memory}
    \acroplural{dram}[DRAMs]{Dynamic Random Access Memories}
    \acro{dsp}[DSP]{Digital Signal Processor}
    \acro{ecpu}[eCPU]{embedded CPU}
    \acro{emem}[eMEM]{embedded Memory}
    \acro{envm}[ENMV]{Embedded Non-Volatile Memory}
    \acroplural{envm}[ENMVs]{Embedded Non-Volatile Memories}
    \acro{fft}[FFT]{Fast Fourier Transform}
    \acro{fll}[FLL]{Frequency-Locked Loop}
    \acro{fpga}[FPGA]{Field-Programmable Gate Array}
    \acro{fsm}[FSM]{Finite State Machine}
    \acro{gcc}[GCC]{GNU Compiler Collection}
    \acro{gemm}[GEMM]{General Matrix Multiplication}
    \acro{gpr}[GPR]{General-Purpose Register}
    \acro{hdl}[HDL]{Hardware Description Language}
    \acro{ic}[IC]{Integrated Circuit}
    \acro{imc}[IMC]{In-Memory Computing}
    \acro{iot}[IoT]{Internet of Things}
    \acro{isa}[ISA]{Instruction Set Architecture}
    \acro{mac}[MAC]{Multiply-and-Accumulate}
    \acro{mcu}[MCU]{Microcontroller Unit}
    \acro{ml}[ML]{Machine Learning}
    \acro{mram}[MRAM]{Magnetoresistive Random Access Memory}
    \acroplural{mram}[MRAMs]{Magnetoresistive Random Access Memories}
    \acro{nmc}[NMC]{Near-Memory Computing}
    \acro{pcm}[PCM]{Phase-Change Memory}
    \acroplural{pcm}[PCMs]{Phase-Change Memories}
    \acro{pe}[PE]{Processing Element}
    \acro{pim}[PIM]{Processing-In-Memory}
    \acro{pvt}[PVT]{Process, Voltage, and Temperature}
    \acro{relu}[ReLU]{Rectified Linear Unit}
    \acro{risc}[RISC]{Reduced Instruction Set Computer}
    \acro{rom}[ROM]{Read-Only Memory}
    \acroplural{rom}[ROMs]{Read-Only Memories}
    \acro{rram}[RRAM]{Resistive Random Access Memory}
    \acroplural{rram}[RRAMs]{Resistive Random Access Memories}
    \acro{rtl}[RTL]{Register Transfer Level}
    \acro{rvv}[RVV]{RISC-V Vector Extension}
    \acro{sew}[SEW]{Selected Element Width}
    \acro{simd}[SIMD]{Single Instruction Multiple Data}
    \acro{soc}[SoC]{System-on-Chip}
    \acroplural{soc}[SoCs]{Systems-on-Chip}
    \acro{sram}[SRAM]{Static Random-Access Memory}
    \acroplural{sram}[SRAMs]{Static Random-Access Memories}
    \acro{vcd}[VCD]{Value Change Dump}
    \acro{vpu}[VPU]{Vector Processing Unit}
    \acro{vrf}[VRF]{Vector Register File}
    \acro{x-heep}[X-HEEP]{eXtendable Heterogeneous Energy-efficient Platform}
    \acro{flsl}[FLSL]{feedforward leakage suppression logic}
    \acro{fdsoi}[FDSOI]{fully depleted silicon-on-insulator}
    \acro{dlsl}[DLSL]{dynamic leakage suppression logic}
    \acro{hvt}[HVT]{high voltage threshold}
    \acro{uhvt}[UHVT]{ultra-high voltage threshold}
    \acro{ehvt}[EHVT]{extreme high voltage threshold}
    \acro{lvt}[LVT]{low voltage threshold}
    \acro{rbb}[RBB]{reverse body-bias}
    \acro{fbb}[FBB]{forward body-bias}
    \acro{ao}[AO]{always-on}
    \acro{pvt}[PVT]{process, voltage, and temperature}
    \acro{ldp}[LDP]{leakage-power delay product}
     \acro{aes}[AES]{advanced encryption standard}
    
\end{acronym}

\bibliographystyle{IEEEtran}
\bibliography{bibliography.bib}

\vfill

\end{document}